%% file: main.tex
\documentclass[3p, twocolumn]{elsarticle}

\usepackage[english]{babel}
\usepackage{pifont}
\usepackage{booktabs} 
\usepackage{etoolbox}
\usepackage[figurename=Fig.,font={small},labelfont={bf,up}]{caption}
\usepackage{subcaption}
\usepackage{balance}
\usepackage{graphicx}
\usepackage{amsmath}
\usepackage{array}
\usepackage{multirow}

\usepackage[ruled,vlined]{algorithm2e}
\usepackage{tablefootnote}

\usepackage{hyperref}
\usepackage{lineno}
\usepackage{tabularx}
\modulolinenumbers[5]

\journal{Computer Networks}

\usepackage{color, colortbl}
\usepackage{longtable}
\usepackage{xcolor}
\usepackage{soul}
\usepackage{float}
\usepackage{comment}
\usepackage{url}
\usepackage{wrapfig}
\usepackage{multirow}
\usepackage{textcomp}
\usepackage{scrextend}
\usepackage{eurosym}
\usepackage{lipsum}
\usepackage{cleveref}

\biboptions{numbers,sort&compress}

\newcommand{\squishlist}{
 \begin{list}{$\bullet$}
  { \setlength{\itemsep}{0pt}
     \setlength{\parsep}{3pt}
     \setlength{\topsep}{3pt}
     \setlength{\partopsep}{0pt}
     \setlength{\leftmargin}{1.5em}
     \setlength{\labelwidth}{1em}
     \setlength{\labelsep}{0.5em} } }

\newcommand{\squishlisttwo}{
 \begin{list}{$\bullet$}
  { \setlength{\itemsep}{0pt}
     \setlength{\parsep}{0pt}
    \setlength{\topsep}{0pt}
    \setlength{\partopsep}{0pt}
    \setlength{\leftmargin}{2em}
    \setlength{\labelwidth}{1.5em}
    \setlength{\labelsep}{0.5em} } }

\newcommand{\squishend}{
  \end{list}  }

\newbool{showComments}
\boolfalse{showComments}

\ifbool{showComments}{%
\newcommand{\zhenyu}[1]{\sethlcolor{orange}\hl{[ZY: #1]}}
\newcommand{\fy}[1]{\sethlcolor{yellow}\hl{[Furong: #1]}}
\newcommand{\da}[1]{\sethlcolor{olive}\hl{[Davide: #1]}}
\newcommand{\gp}[1]{\sethlcolor{cyan}\hl{[Giovanni: #1]}}
\newcommand{\pd}[1]{\sethlcolor{green}\hl{[Pesa: #1]}}
}{

\newcommand{\zhenyu}[1]{} 
\newcommand{\af}[1]{} 
\newcommand{\fy}[1]{}
\newcommand{\da}[1]{}
\newcommand{\gp}[1]{}
\newcommand{\pd}[1]{}
}

\newcommand{\algname}[1]{BBRv2+}
\newcommand{\signame}[1]{delay information}

\begin{document}

\begin{frontmatter}


\title{BBRv2+: Towards Balancing Aggressiveness and Fairness with Delay-based Bandwidth Probing}



\author[ICT,UCAS,SU]{Furong Yang}
\ead{yangfurong@ict.ac.cn}

\author[ICT]{Qinghua Wu}
\ead{wuqinghua@ict.ac.cn}

\author[ICT]{Zhenyu Li\corref{mycorrespondingauthor}}
\cortext[mycorrespondingauthor]{Corresponding author}
\ead{zyli@ict.ac.cn}

\author[ALIBABA]{Yanmei Liu}
\ead{miaoji.lym@alibaba-inc.com}

\author[UNIBO,UCLA]{Giovanni Pau}
\ead{giovanni.pau@unibo.it}

\author[CNIC]{Gaogang Xie}
\ead{xie@cnic.cn}

\address[ICT]{Institute of Computing Technology, Chinese Academy of Sciences, China}
\address[UCAS]{University of Chinese Academy of Sciences, China}
\address[SU]{Sorbonne University, France}
\address[ALIBABA]{Alibaba Group, China}
\address[UNIBO]{University of Bologna, Italy}
\address[UCLA] {University of California, Los Angeles, USA}
\address[CNIC]{Computer Network Information Center, Chinese Academy of Sciences, China}

\input{sections/abstract}


\end{frontmatter}

\section{Introduction}
\input{sections/introduction}

\section{Background: an overview of BBR and BBRv2}
\input{sections/background}

\section{A Deep Dive into BBRv2}
\input{sections/measurement}

\section{Design and implementation of BBRv2+}
\input{sections/motivation-arch}

\section{Evaluation of BBRv2+}
\input{sections/BBRv2p-evaluation}

\section{Related work}
\input{sections/relatedwork}

\section{Conclusion}
\input{sections/conclusion}
\bibliography{mybibfile}

\end{document}

%% file: sections/abstract.tex
\begin{abstract}
BBRv2, proposed by Google, aims at addressing BBR's shortcomings of unfairness against loss-based congestion control algorithms (CCAs) and excessive retransmissions in shallow-buffered networks. In this paper, we first comprehensively study BBRv2's performance under various network conditions and show that  BBRv2 mitigates the shortcomings of BBR. Nevertheless, BBRv2's benefits come with several costs, including the slow responsiveness to bandwidth dynamics as well as the low resilience to random losses. We then propose \algname{} to address BBRv2's performance issues without sacrificing its advantages over BBR. To this end, \algname{} incorporates \signame{} into its path model, which cautiously guides the aggressiveness of its bandwidth probing to not reduce its fairness against loss-based CCAs. \algname{} also integrates mechanisms for improved resilience to random losses as well as network jitters. Extensive experiments demonstrate the effectiveness of \algname{}. Especially, it achieves 25\% higher throughput and comparable queuing delay in comparison with BBRv2 in high-mobility network scenarios. 
\end{abstract}

%% file: sections/introduction.tex
Congestion control has been one of the active research topics in computer networks since it was introduced in the 1980s~\cite{Jacobson1995CongestionAA}. More than three decades of research on congestion control have brought us a plethora of congestion control algorithms (CCAs) and TCP variants, aiming at efficient utilization of available bandwidth while fairly sharing the bottleneck bandwidth among multiple flows. For instance, Linux kernel alone has more than 15 different CCAs~\cite{abbasloo_classic_2020}. While we see many recent proposals on learning-based CCAs (e.g. Remy~\cite{winstein_tcp_nodate}, Aurora~\cite{Jay2019ADR}, PCC-Vivace~\cite{Dong2018PCCVO}, Indigo~\cite{Yan2018PantheonTT}, Orca~\cite{abbasloo_classic_2020}), the wildly deployed CCAs today are still classic ones (e.g. Cubic~\cite{rhee_cubic_2018}, BBR~\cite{Cardwell2016BBRCC}). 

In this paper, we focus on BBR and its upgrade, BBRv2, as BBR has been used by 22\% of the Alexa Top 20K websites~\cite{mishra_great_2019} and BBRv2 will likely replace BBR in the near future\footnote{As of March 2021, Google has finished roll-out of BBRv2 for internal TCP traffic, and tuning performance to enable roll-out for external traffic~\cite{yeganeh_bbr_110}.}. BBR is a rate-based CCA that sets its sending rate based on the measured bottleneck bandwidth (\textit{BtlBW}) and round trip propagation time (\textit{RTprop}). That said, instead of reacting to congestion signals such as losses or delay dynamics, BBR tries to actively operate at Kleinrock's optimal operating point~\cite{Kleinrock1979PowerAD} to maximize throughput without incurring standing queues at a bottleneck link. The previous empirical studies~\cite{cao_when_2019, scholz_towards_2018, hock_experimental_2017, ware_modeling_2019, kumar_tcp_2019, ma_fairness_2017} have disclosed several shortcomings of BBR: \textbf{(1)} it causes excessive retransmissions in shallow-buffered networks; \textbf{(2)} it is not fair when competing with flows using loss-based CCAs (e.g. Cubic) and BBR flows with different Round Trip Times (RTTs); \textbf{(3)} its performance degrades when network jitters are high. 

To address these issues in BBR, Google proposed BBRv2~\cite{cardwell_bbr_104} that inherits most of the design principles from BBR, while reacts to losses and Explicit Congestion Notification (ECN) marks for better co-existence with loss-based CCAs and being less aggressive in shallow-buffered networks. Given that BBRv2 may eventually replace BBR, understanding how BBRv2 actually performs is of great importance for improved performance and fairness. We have also seen several studies~\cite{gomez_performance_2020, cardwell_bbr_105, nandagiri_bbrvl_2020, song_understanding_2021, kfoury_emulation-based_2020} on measuring BBRv2, which have shown that, in comparison with BBR, BBRv2 improves the inter-protocol fairness against loss-based CCAs and reduce retransmissions in shallow-buffered networks. Nevertheless, we found in this paper, these improvements come with several costs, including the low resilience to random packet losses and the slow responsiveness to bandwidth dynamics. 

In this paper, we first evaluate BBRv2 in various network conditions with an emphasis on the reasons behind the observed performance issues. Our key observations from the empirical study of BBRv2 are as follows: 
\squishlist
    \item Due to its conservative strategies in bandwidth probing and inflight cap estimation, BBRv2 achieves better inter-protocol fairness against loss-based CCAs in shallow-buffered networks than BBR. On the other hand, these strategies also make BBRv2 slightly less competitive than BBR in terms of throughput under moderate buffers. BBRv2 also improves RTT fairness among flows, compared with BBR.
    
    \item In shallow-buffered networks, retransmissions of BBRv2 are significantly reduced compared with that of BBR. However, the throughput of BBRv2 is 13\%$\sim$16\% lower than that of BBR under shallow buffers, as BBRv2 limits its inflight size to about 0.85$\times$ BDP for most of the time in these networks.
    
    \item BBRv2 is less resilient to random losses than BBR. Interestingly, we find that carefully tuning the loss threshold parameter in BBRv2 according to bottleneck buffer sizes can enhance BBRv2's loss resilience without sacrificing its advantages in retransmission and fairness.
    
    \item BBRv2 is less responsive to bandwidth dynamics than BBR, which leads to low bandwidth utilization and high queuing delay in networks with bandwidth dynamics. The long bandwidth probing interval and the long expiry time of bottleneck bandwidth estimation are the two major contributors.
    
    \item Like BBR, BBRv2's performance still suffers from congestion window (\textit{cwnd}) exhaustion in high-jitter networks that are not rare in wireless scenarios~\cite{zhang_will_2018, chitimalla_5g_2017, kumar_tcp_2019, beshay_link-coupled_2017, wang_active-passive_2019}.
\squishend

The results of our empirical study of BBRv2 raise one question to us: \textit{are we able to improve BBRv2's performance while keeping its advantages in retransmission and fairness? If so, how do we achieve this goal?}

Compared with BBR, BBRv2's shortcomings lie in the lower loss resilience and slower responsiveness to bandwidth dynamics. On one hand, the issue regarding loss resilience can be mitigated by carefully tuning the loss threshold parameter in BBRv2. On the other hand, we can increase the aggressiveness of BBRv2 in bandwidth probing to improve its responsiveness to bandwidth dynamics. But, this aggressiveness needs to be cautiously guided, as blindly behaving aggressively may cause unfairness against loss-based CCAs like BBR. Currently, the aggressiveness of bandwidth probing in BBR and BBRv2 is either hard-coded or pre-configured according to designers' experience without the perception of the network environment. This kind of unguided aggressiveness may make the bandwidth probing be either over-aggressive (like BBR) or over-conservative (like BBRv2) in certain environments. Thus, the feedback from the network environment needs to be considered in BBRv2's bandwidth probing strategy to guide the aggressiveness of bandwidth probing.

To address the above gap, we propose \algname{}. Firstly, \algname{} integrates \signame{} into its path model, which serves as the feedback to guide its aggressiveness in bandwidth probing. Secondly, to utilize the \signame{} to guide the aggressiveness of \algname{}'s bandwidth probing, the state-machine of BBRv2 is partially redesigned. In doing so, \algname{} balances between the aggressiveness in probing for more bandwidth and the fairness against loss-based CCAs. Thirdly, to avoid being suppressed when co-existing with loss-based CCAs in deep-buffered networks because of using the \signame{}, \algname{} incorporates a dual-mode mechanism, where it switches to use BBRv2's state-machine if no RTT sample approaching \textit{RTprop} is observed for a long time period and returns back to use the redesigned state-machine if it constantly observes RTT samples approaching \textit{RTprop}. Finally, as an optimization, \algname{} addresses the \textit{cwnd} exhaustion problem in high-jitter networks by compensating its estimated Bandwidth Delay Product (BDP) according to observed jitters; the compensation mechanism allows the estimated BDP to be close to the actual BDP, and can also be applied to BBR and BBRv2.

Extensive experiments based on both Mininet and Mahimahi with real-world traces show that compared with BBRv2, \algname{} succeeds to balance the aggressiveness of bandwidth probing and the fairness against loss-based CCAs, improves the resilience to network jitters, and, particularly, achieves 25\% higher throughput and comparable queuing delay in high-mobility scenarios where the bandwidth is very dynamic.


To summarize, the contributions of this paper are three-fold: \textbf{(1)} a deep dive into BBRv2 that reveals its pros and cons, compared with BBR; \textbf{(2)} \algname{} that addresses the shortcomings of BBRv2 while barely sacrificing BBRv2's advantages; \textbf{(3)} extensive experiments demonstrating that \algname{} meets its design goals. We open-source \algname{} to the research community for further test and improvement~\cite{furongYangfurongBBRv2plus2021}.

The remainder of this paper is organized as follows. We first give an overview of BBR and BBRv2 in \S\ref{sec:background}. Then, a deep dive into BBRv2, which motivates the design of \algname{}, is presented in \S\ref{sec:measurement}. Next, the design and implementation of \algname{} is described in \S\ref{sec:design}, and the evaluation of \algname{} is shown in \S\ref{sec:eva}. After that, we present the related work in \S\ref{sec:related_works}. Finally, the paper is concluded in \S\ref{sec:conclusion}.

%% file: sections/background.tex
\label{sec:background}

BBR aims at maximizing throughput while keeping the lowest latency; it requires accurate measurements of both \textit{BtlBW} and \textit{RTprop}. Since these two variables cannot be measured simultaneously, BBR introduces a state-machine-based method that alternatively estimates \textit{BtlBW} and \textit{RTprop}.



\begin{figure}
    \centering
    \includegraphics[width=0.8\linewidth]{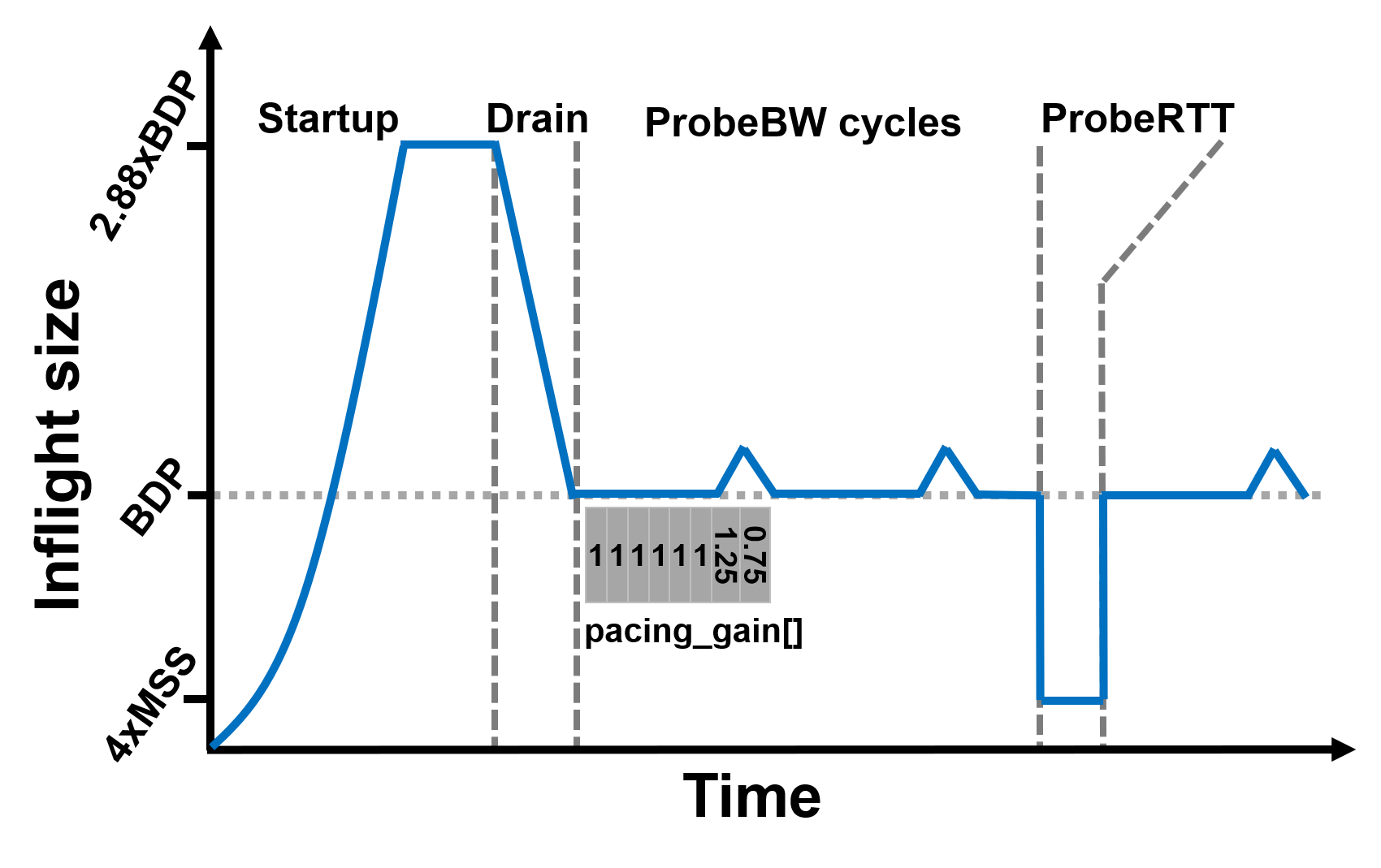}
    \caption{Illustration of BBR life-cycle}
    \label{fig:BBR_lifecycle}
\end{figure}

\begin{figure}
    \centering
    \includegraphics[width=0.9\linewidth]{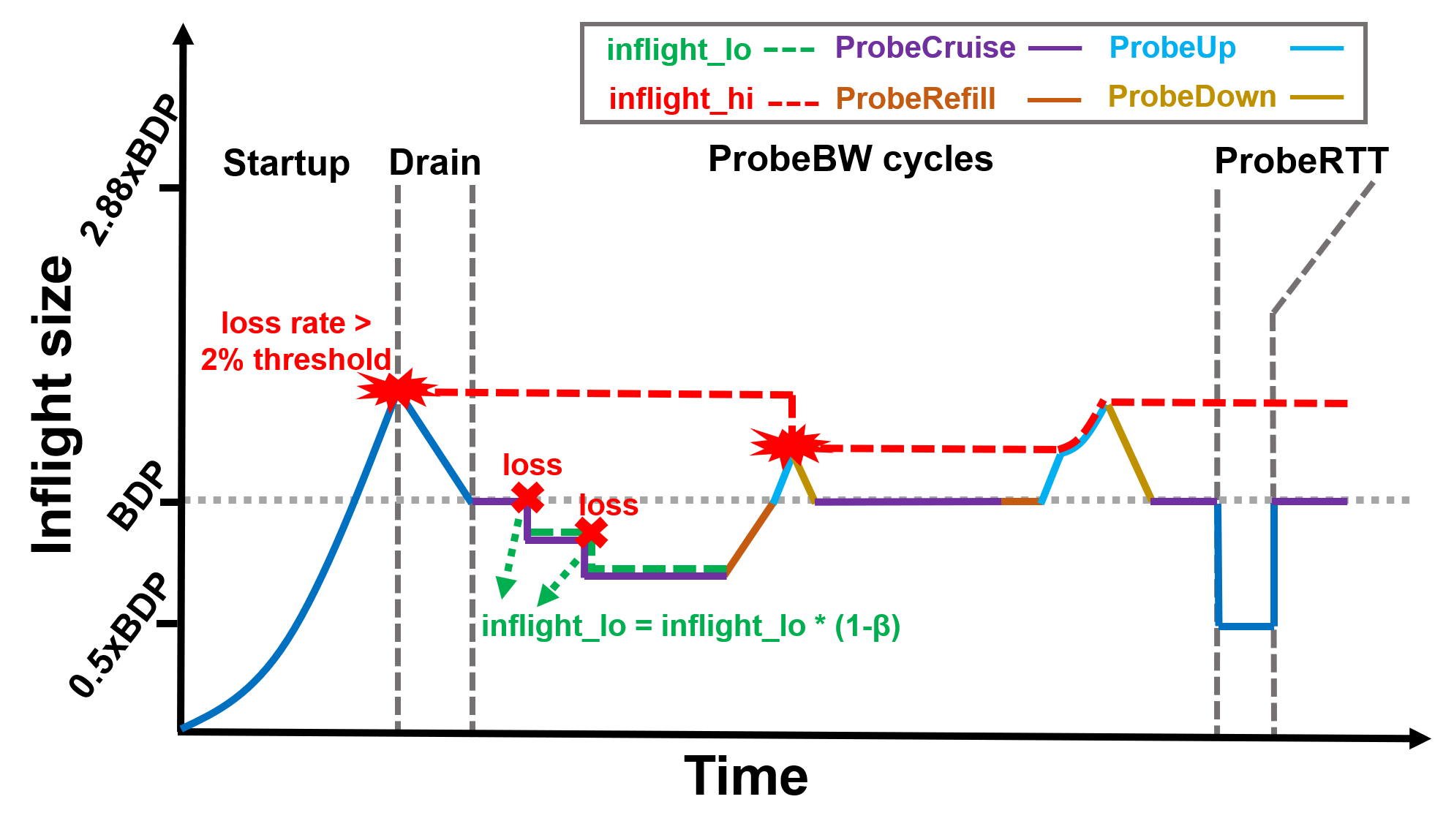}
    \caption{Illustration of BBRv2 life-cycle}
    \label{fig:BBRv2_lifecycle}
\end{figure}

As illustrated in Fig.~\ref{fig:BBR_lifecycle}, there are four states in the BBR life-cycle. BBR uses \emph{pacing\_gain} to control the sending behavior---to probe for more bandwidth, to drain the queue at the bottleneck link, or to cruise at the speed of \textit{BtlBW}. Firstly, BBR starts the Startup state which exponentially increases the inflight size and sending rate by setting \emph{pacing\_gain} to $\mathrm{2/ln(2)}$. BBR transits into the Drain state when it is in the plateau of \textit{BtlBW} for three RTTs. In the Drain state, BBR reduces \emph{pacing\_gain} to $\mathrm{ln(2)/2}$ to drain the standing queue at the bottleneck link induced during Startup. After the above two stages, BBR has successfully built the path model, with \textit{RTprop} measured at the beginning of Startup and \textit{BtlBW} measured at the end of Startup. After that, BBR switches to a steady phase where BBR alternatively runs in the ProbeBW and ProbeRTT state. During the ProbeBW state, BBR sets \emph{pacing\_gain} to 1.0 to cruise at Kleinrock's optimal point for 6 cycles, then sets \emph{pacing\_gain} to 1.25 to explore more bandwidth for 1 cycle and thereafter sets \emph{pacing\_gain} to 0.75 to drain the possible standing queue for 1 cycle. In the ProbeRTT state, BBR reduces its inflight size to 4$\times$ MSS (Max Segment Size) and waits for $max\{RTT, 200ms\}$ to measure an updated value of \textit{RTprop}.

As observed in numerous previous studies~\cite{cao_when_2019, scholz_towards_2018, hock_experimental_2017, ware_modeling_2019, kumar_tcp_2019}, BBR has two key issues: the unfairness of bandwidth share with loss-based CCAs and the high retransmission rate in shallow-buffered networks. The reasons behind these issues are that BBR is congestion signal agnostic and is over-aggressive when probing for more bandwidth. To mitigate the problems above, Google proposed BBRv2, which inherits most of BBR's design (e.g. the core principle, the overall building blocks, etc.) yet redesigns the ProbeBW state, as illustrated in Fig.~\ref{fig:BBRv2_lifecycle}.

BBRv2 adds measurements of packet loss and DCTCP-style ECN marks~\cite{Alizadeh2010DataCT} for estimating the capacity of a bottleneck link. Specifically, it introduces \emph{inflight\_lo} and \emph{bw\_lo} as the short-term lower bounds of inflight size and sending rate respectively, in order to capture the temporary status of the network path (e.g. cross-traffic takes a share of capacity); it uses \emph{inflight\_hi} as the long-term upper bound of inflight size to reduce the likelihood of packet loss. To avoid recklessly probing for more bandwidth, BBRv2 decomposes BBR's ProbeBW state into four sub-states: ProbeCruise, ProbeRefill, ProbeUp, and ProbeDown.

\noindent\textbf{ProbeCruise:} In ProbeCruise, BBRv2 sets \emph{pacing\_gain} to 1. If any loss or ECN mark occurs, BBRv2 updates \emph{inflight\_lo} and \emph{bw\_lo} to $max\{(1-\beta)\times inflight\_lo, \textit{BtlBW}_{curr}\}$ and $max\{(1-\beta)\times bw\_lo, inflight_{curr}\}$ respectively, where $\textit{BtlBW}_{curr}$ and $inflight_{curr}$ are the current measurements of bandwidth and inflight size.

\noindent\textbf{ProbeRefill:} When BBRv2 has been in ProbeCruise for a period of $T$ ($T$ is determined in ProbeDown), BBRv2 transits to ProbeRefill, by setting \emph{inflight\_lo} and \emph{bw\_lo} to $+\infty$ to refill the ``pipe'' with BDP-sized inflight data, which lasts for one RTT. The goal of this state is to avoid early losses before the capacity is fully utilized in shallow-buffered networks since BBRv2 will accelerate in the following ProbeUp state, which may lead the bottleneck buffer to overflow.

\noindent\textbf{ProbeUp:} During ProbeUp, BBRv2 sets \emph{pacing\_gain} to 1.25 to probe for more available bandwidth. This state ends either when the current loss rate exceeds a pre-defined explicit loss threshold (2\%), (or the ECN mark rate exceeds an ECN threshold), or when the inflight size reaches 1.25$\times$ BDP and at least one \textit{RTprop} has passed. In the former case, BBRv2 sets \emph{inflight\_hi} to the current inflight size.

\noindent\textbf{ProbeDown:} During ProbeDown, BBRv2 drains the potential queue at the bottleneck link by setting \emph{pacing\_gain} to 0.75. BBRv2 also sets the duration ($T$) for the next ProbeCruise state to $min\{rand(2,3), \frac{BDP}{MSS}\times{}RTT\}$ seconds, where $rand(2,3)$ means a number between two and three. The intention of $T$ is to match the interval between loss recovery epochs of Reno for TCP fairness. The ProbeDown state ends when BBRv2 cuts its inflight size below the minimum value between 1$\times$ BDP and $0.85\times {inflight\_hi}$. Thereafter, BBRv2 transits to the next ProbeCruise state.

As the bandwidth probing behaviors of BBRv2 are different from BBR, BBRv2 no longer uses a 10-RTT-windowed \textit{max\_filter} to track the estimation of \textit{BtlBW}, and it rather takes the maximum bandwidth measured in the recent two ProbeBW stages as the estimation of \textit{BtlBW}, which ensures that the bandwidth samples from ProbeUp states are considered.

\vspace{0.5em}
\noindent\textbf{Summary of BBRv2:} The inflight bound mechanism, driven by losses or ECN marks, and the less aggressive bandwidth probing strategy make BBRv2 more conservative than BBR in shallow-buffered networks, which thus mitigates the problems regarding excessive retransmissions and unfairness. However, the two changes can potentially reduce BBRv2's performance under random losses and bandwidth dynamics, because, compared with BBR, BBRv2 probes for bandwidth less frequently, takes more time to expire \textit{BtlBW} estimations, and slows down constantly if the random loss rate exceeds 2\%.

%% file: sections/measurement.tex
\label{sec:measurement}


In this section, we conduct extensive measurements to investigate the improvements and overheads of BBRv2, in comparison with BBR. Our key observations include: \textbf{(1)} BBRv2 improves the inter-protocol fairness and RTT fairness, and also reduces retransmissions in shallow-buffered networks; this observation reaffirms those in the previous studies~\cite{gomez_performance_2020, cardwell_bbr_105, nandagiri_bbrvl_2020, song_understanding_2021, kfoury_emulation-based_2020}; \textbf{(2)} the improvements of BBRv2 come with the cost of the low resilience to random loss and the slow responsiveness to bandwidth dynamics. That said, it fails to achieve a balance between the aggressiveness in probing for more bandwidth and the fairness against loss-based CCAs; \textbf{(3)} like BBR, BBRv2 experiences low throughput in high-jitter networks because of underestimation of BDP.

\subsection{Methodology}
\label{sec:measurement:method}

\begin{figure}[htb]
    \centering
    \includegraphics[width=0.8\linewidth]{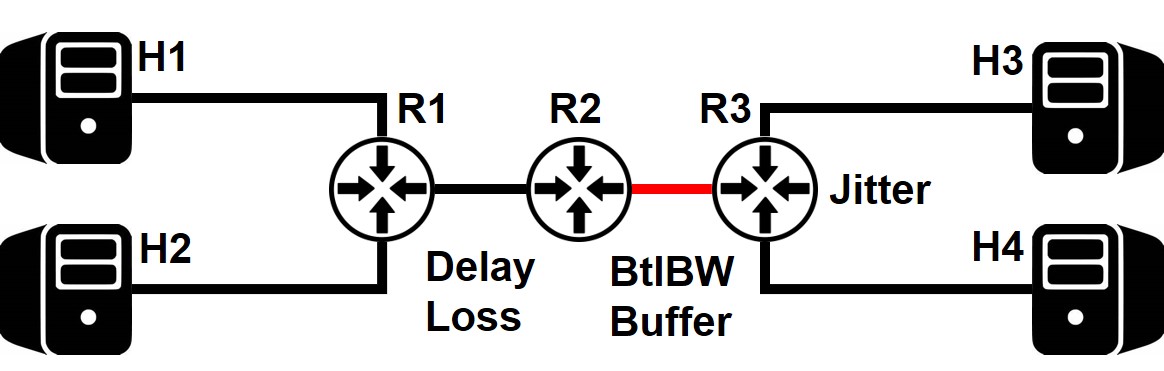}
    \caption{Mininet testbed}
    \label{fig:mininet_topo}
\end{figure}

We utilize Mininet~\cite{noauthor_mininet_nodate} to build an emulation-based testbed. The testbed, whose topology is shown in Fig.~\ref{fig:mininet_topo}, was run on a server with 8 Intel Xeon Platinum cores and 32GB of memory. The operating system is Ubuntu 18.04.5 with BBR and BBRv2~\cite{bbr2_kernel} installed. Linux \textit{tc-netem}~\cite{noauthor_tc-netem8_nodate} is used to emulate different network conditions (e.g. router buffer size, link speed, RTT, random loss rate, jitter). \textit{Iperf3}~\cite{noauthor_esnetiperf_nodate} generates TCP traffic between senders and receivers. During the transmission, various performance metrics (e.g. RTT, throughput, retransmissions, inflight bytes) are measured by \textit{tcpdump}~\cite{noauthor_tcpdumplibpcap_nodate} and \textit{tcptrace}~\cite{noauthor_tcptrace1_nodate}. Moreover, a set of internal variables (e.g. \textit{cwnd}, \textit{pacing\_rate}, \textit{RTprop}, \textit{BtlBW}) in BBR and BBRv2 are reported by Linux kernel module and the backlog information of the standing queue in bottleneck routers (R2 and R3) is reported by \textit{tc}~\cite{noauthor_tc8_nodate}. Each set of experiments is repeated five times and the average results are reported.

\subsection{Fairness}
\label{sec:measurement:fairness}


We first evaluate the fairness of BBRv2. Two types of fairness are investigated: the inter-protocol fairness against loss-based CCAs and RTT fairness. In this set of experiments, two flows start simultaneously, one from H1 to H3 and the other from H2 to H4, and last for three minutes for the convergence of throughput. The bottleneck bandwidth is fixed at 40Mbps without network jitters or random losses.

We use Jain's fairness index~\cite{rfc5166} ($\mathcal{F}$) of the two flows as the metrics of fairness, calculated according to Eq.~\ref{eq:jain_index}, where $T_i$ is the average throughput of the $i$-th flow. 

\begin{equation}
    \centering
    \mathcal{F} = \frac{(T_{1} + T_{2})^2}{2*(T_{1}^2 + T_{2}^2)}
    \label{eq:jain_index}
\end{equation}

$\mathcal{F} = 1$ indicates the maximum fairness where two flows have the same average throughput, and $\mathcal{F} = 0.5$ represents that one flow's throughput is zero and the fairness is minimized. As the size of the bottleneck buffer also impacts fairness results, we varied the buffer size to study their relationship.

\subsubsection{Inter-protocol fairness}
\label{sec:measurement:inter_fair}


In the experiment, the flow from H1 to H3 uses either BBR or BBRv2, and that from H2 to H4 uses Cubic, which is the default CCA in Linux and MacOS. The RTTs of the two paths of the two flows are set to 40ms.

\begin{figure}
    \centering
    \begin{minipage}{1\linewidth}
        \begin{minipage}{0.48\linewidth}
            \centering
            \includegraphics[width=1\linewidth]{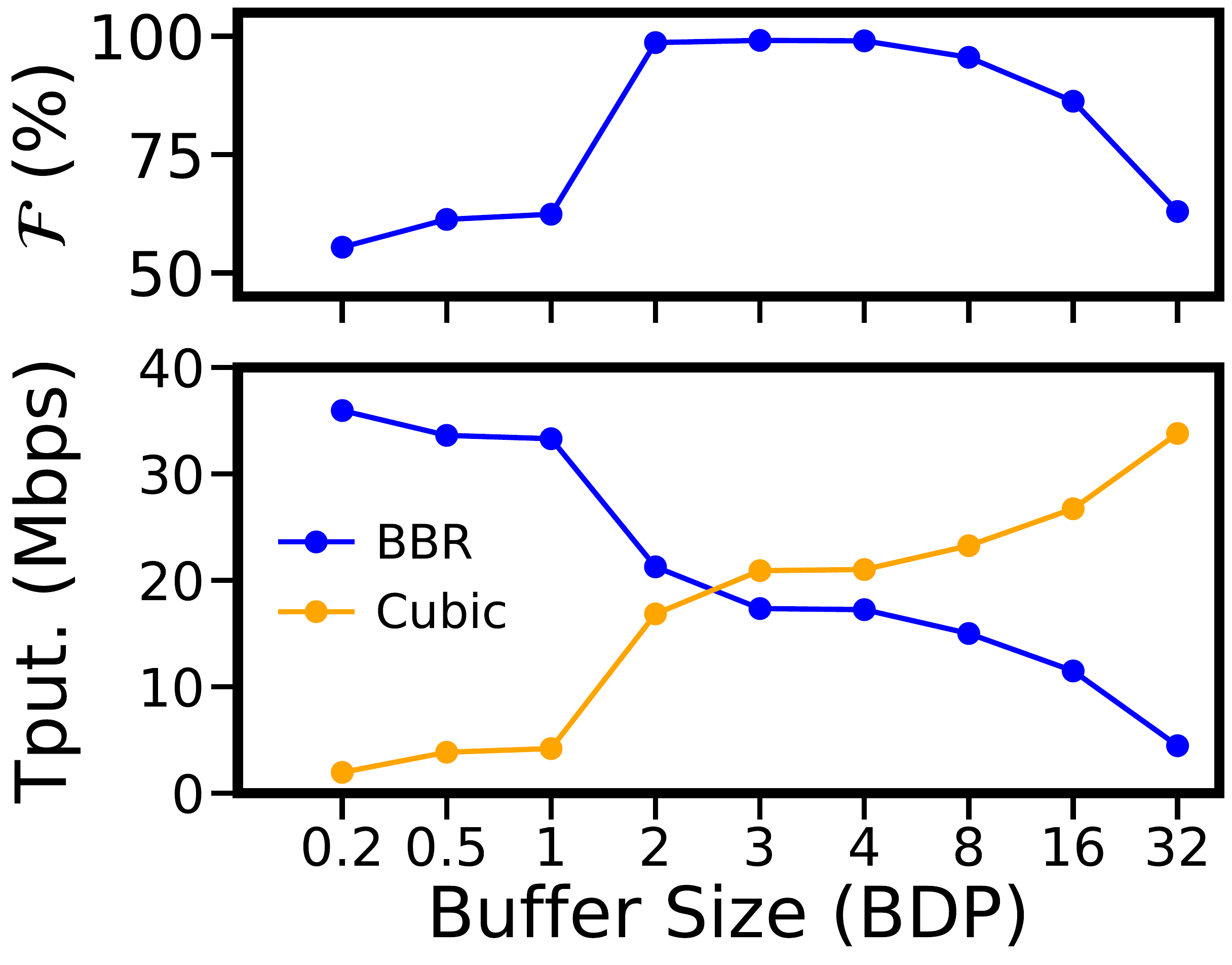}
            \subcaption{BBR vs Cubic}
            \label{fig:cc_fairness_bbr}
        \end{minipage}
        \hfill
        \begin{minipage}{0.48\linewidth}
            \centering
            \includegraphics[width=1\linewidth]{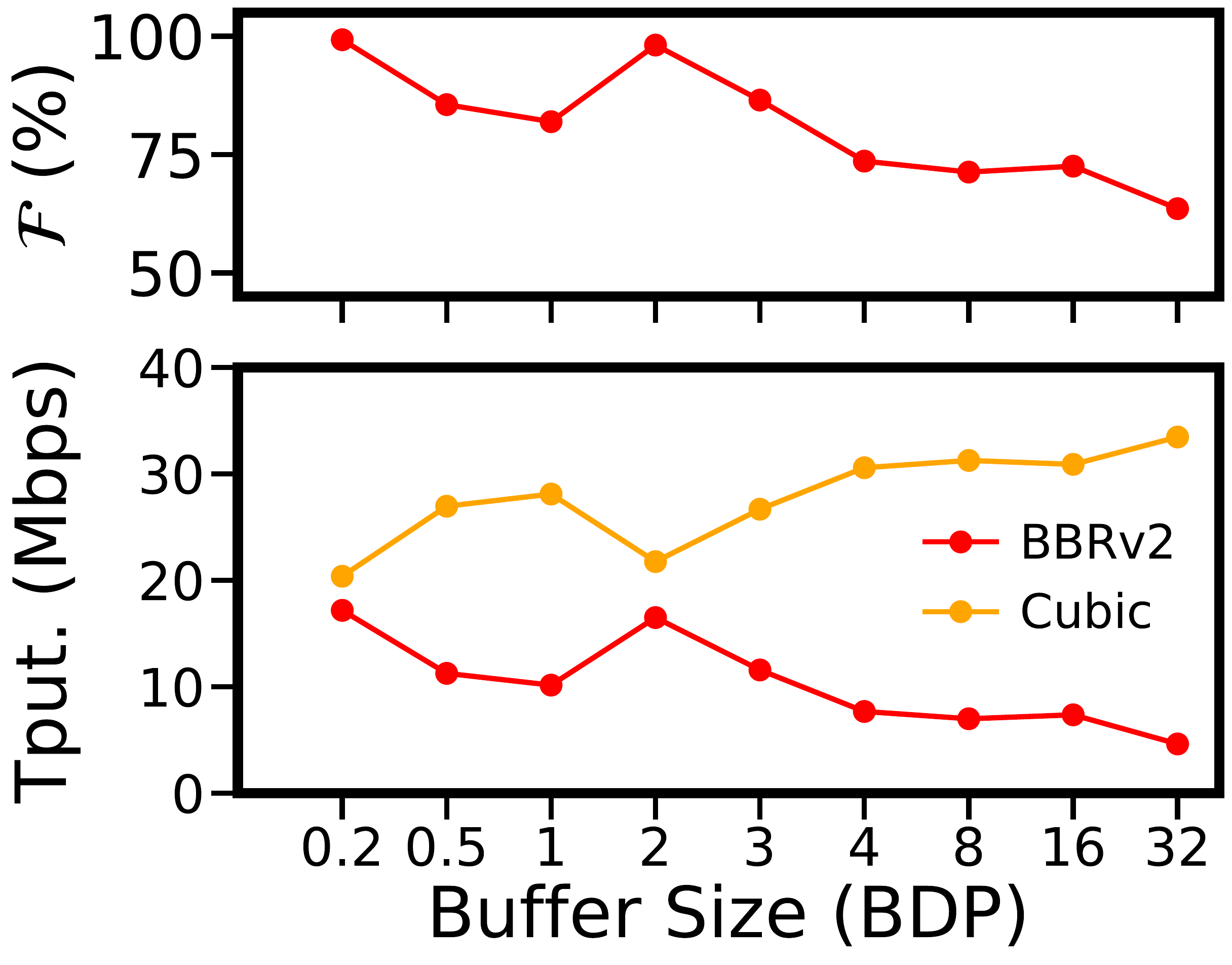}
            \subcaption{BBRv2 vs Cubic}
            \label{fig:cc_fairness_bbr2}
        \end{minipage}
        \caption{Inter-protocol fairness of BBR/BBRv2 under different buffer sizes.}
        \label{fig:cc_fairness}
    \end{minipage}
\end{figure}

Fig.~\ref{fig:cc_fairness} shows the inter-protocol fairness results for both BBR and BBRv2. We can observe that compared with BBR, BBRv2 significantly improves Jain's fairness index when the buffer is shallow (i.e., less than 2$\times$ BDP). This is due to the fact that BBRv2 reacts to losses caused by buffer overflow and bounds the inflight size by using \emph{inflight\_hi} and \emph{inflight\_lo}. When the bottleneck buffer becomes deeper, Cubic obtains more bandwidth than  BBR or BBRv2. This is because that the inflight size of both BBR and BBRv2 is limited by about 2$\times$ BDP, while Cubic's inflight size can go beyond this value under deep buffers. As the two flows experience similar RTTs, a larger inflight size means higher throughput. We also note that BBRv2 is less competitive than BBR under moderate buffers. The reason lies in that BBRv2 is more conservative in bandwidth probing and inflight cap estimation.

\subsubsection{RTT fairness}
\label{sec:measurement:rtt_fair}

\begin{figure}[htb]
    \centering
    \begin{minipage}{1\linewidth}
        \begin{minipage}[t]{0.48\linewidth}
            \centering
            \includegraphics[width=1\linewidth]{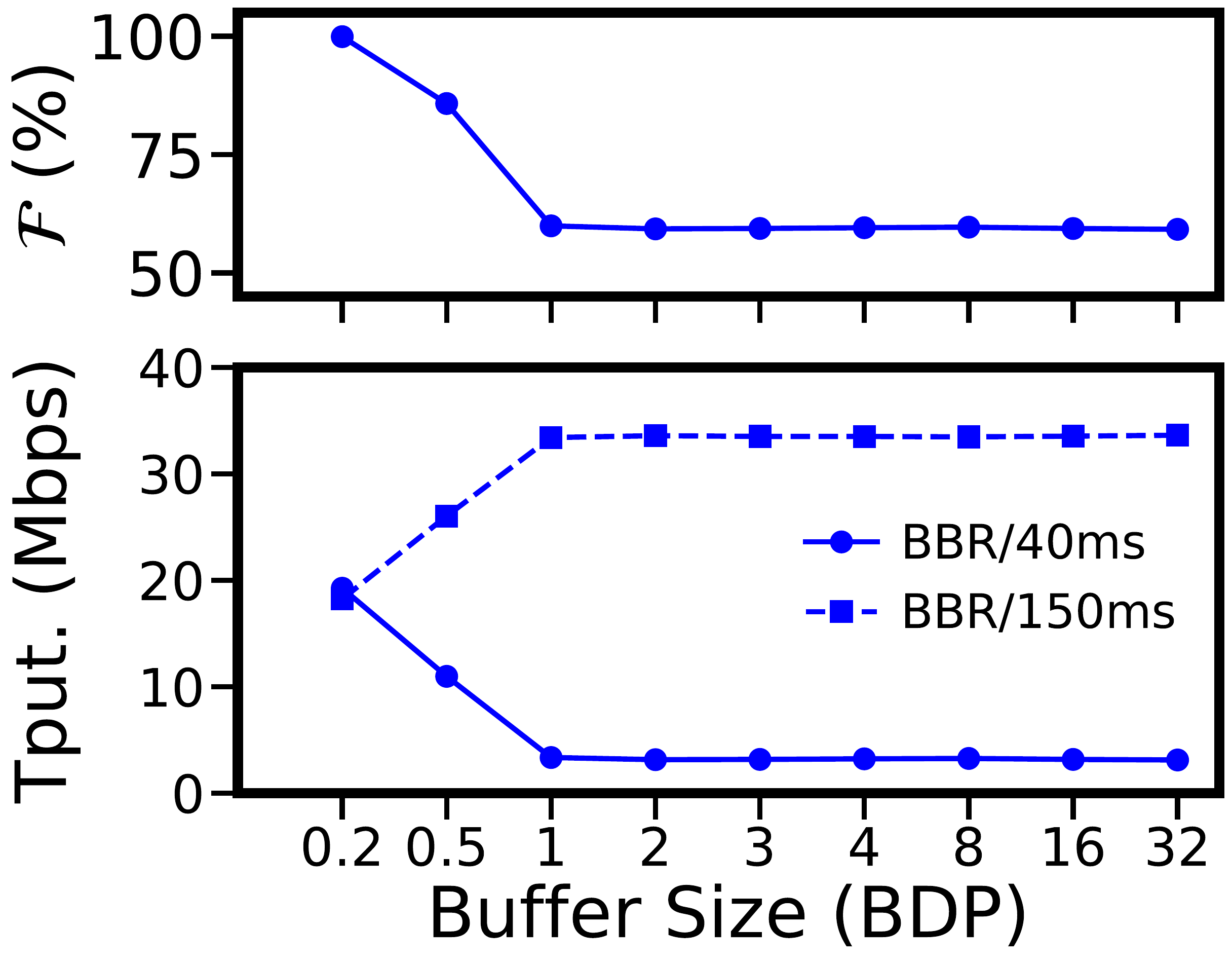}
            \subcaption{BBR}
            \label{fig:rtt_fairness_bbr}
        \end{minipage}
        \hfill
        \begin{minipage}[t]{0.48\linewidth}
            \centering
            \includegraphics[width=1\linewidth]{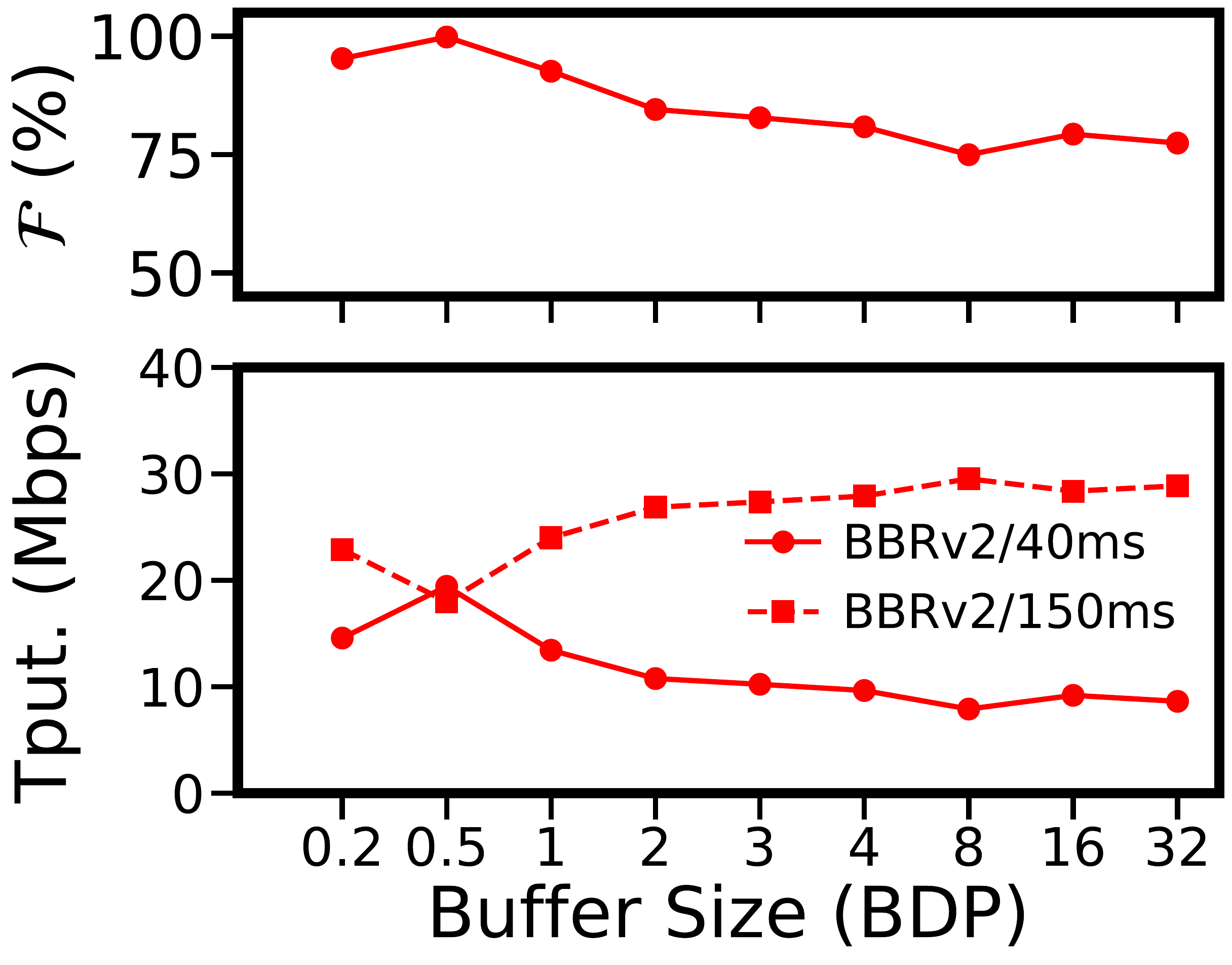}
            \subcaption{BBRv2}
            \label{fig:rtt_fairness_bbr2}
        \end{minipage}
        \caption{RTT fairness of BBR/BBRv2 under different buffer sizes.}
        \label{fig:rtt_fairness}
    \end{minipage}
\end{figure}

In this experiment, both flows use the same CCA, either BBR or BBRv2. The path between H1 and H3 has an RTT of 40ms and that between H2 and H4 has an RTT of 150ms. Fig.~\ref{fig:rtt_fairness} shows the RTT fairness results of BBR and BBRv2, where we set the buffer size to $x$ times of the BDP of the path between H2 and H4.

When the buffer is quite shallow (i.e., 0.2$\times$ BDP), BBR has good fairness. When the buffer size becomes larger, the BBR flow with longer RTT gradually occupies all the bandwidth and starves the flow with shorter RTT. The reason for the poor RTT fairness of BBR is well documented by the previous studies~\cite{hock_experimental_2017, ma_fairness_2017}. The bandwidth probing of BBR leads the aggregated sending rate of two flows to exceed bottleneck bandwidth, thus, forming a persistent queue at the bottleneck link. As the inflight cap of BBR is proportional to \textit{RTprop}, the flow with longer RTT pours more data into the bottleneck buffer, thus, leading to a larger share of the bottleneck link's capacity. This problem is not severe under shallow buffers, as excess packets are mostly dropped instead of forming a persistent queue.

Compared with BBR, BBRv2 has better RTT fairness, especially under deep buffers. The likely reason is three-fold. First, when the buffer size is moderate, losses are triggered due to buffer overflow, and then both flows reduce their inflight size proportional to BDP. As the flow with longer RTT has a larger BDP, it reduces its inflight size more than the flow with shorter RTT. Second, when BBRv2 flows are cruising at the speed of \textit{BtlBW}, they always try to leave headroom\footnote{BBRv2 always limits its inflight size below $0.85\times$ \emph{inflight\_hi} to leave headroom for faster throughput convergence with other flows if there is any.} for other flows to explore the bandwidth. Third, BBRv2 enters the ProbeRTT state more often than BBR, thus, leading BBRv2 flows to yield occupied capacity more frequently.

\begin{figure}[thb]
    \begin{minipage}{1\linewidth}
        \begin{minipage}[t]{0.48\linewidth}
            \centering
            \includegraphics[width=1\linewidth]{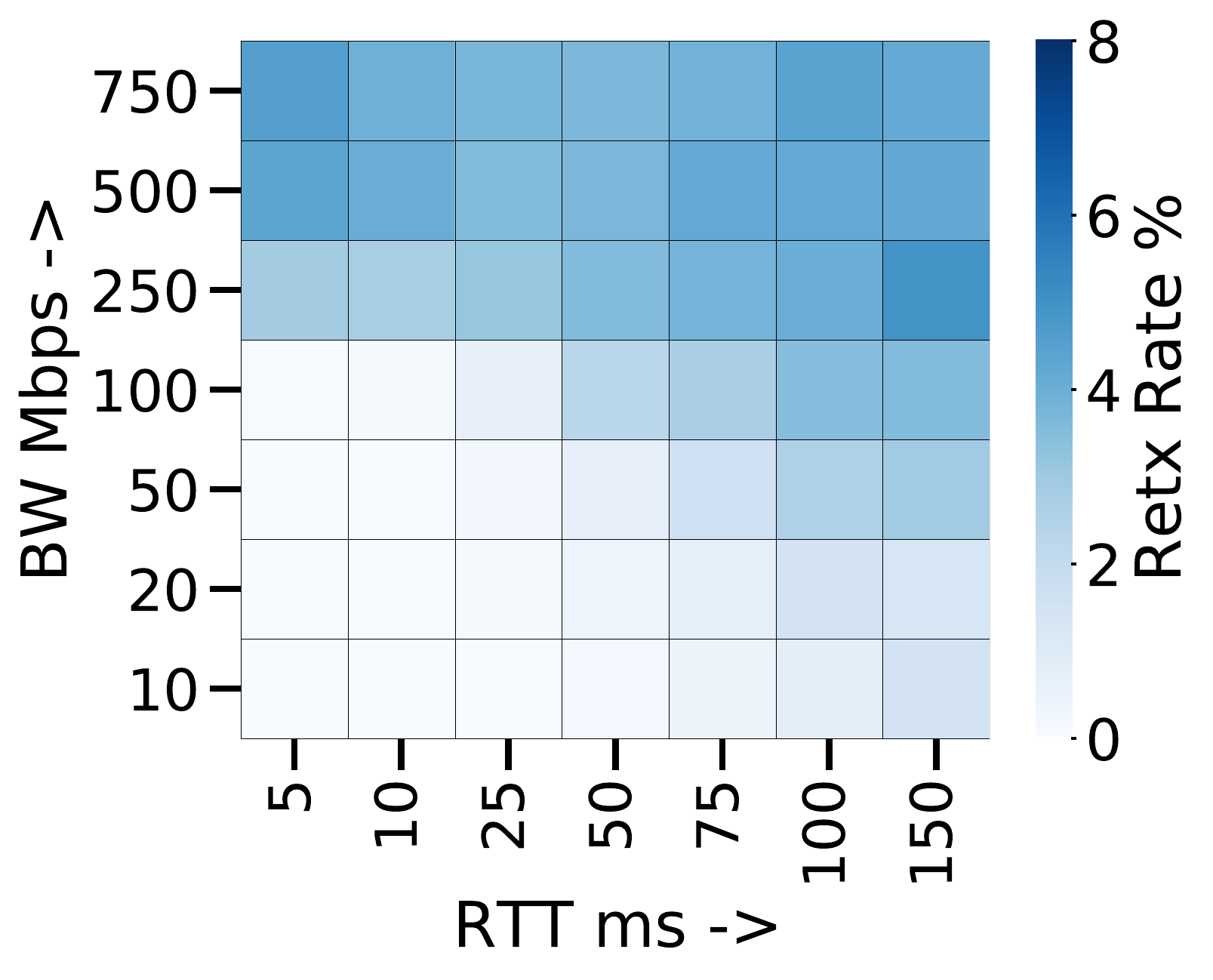}
            \subcaption{BBR}
            \label{fig:retx_bbr}
        \end{minipage}
        \hfill
        \begin{minipage}[t]{0.48\linewidth}
            \centering
            \includegraphics[width=1\linewidth]{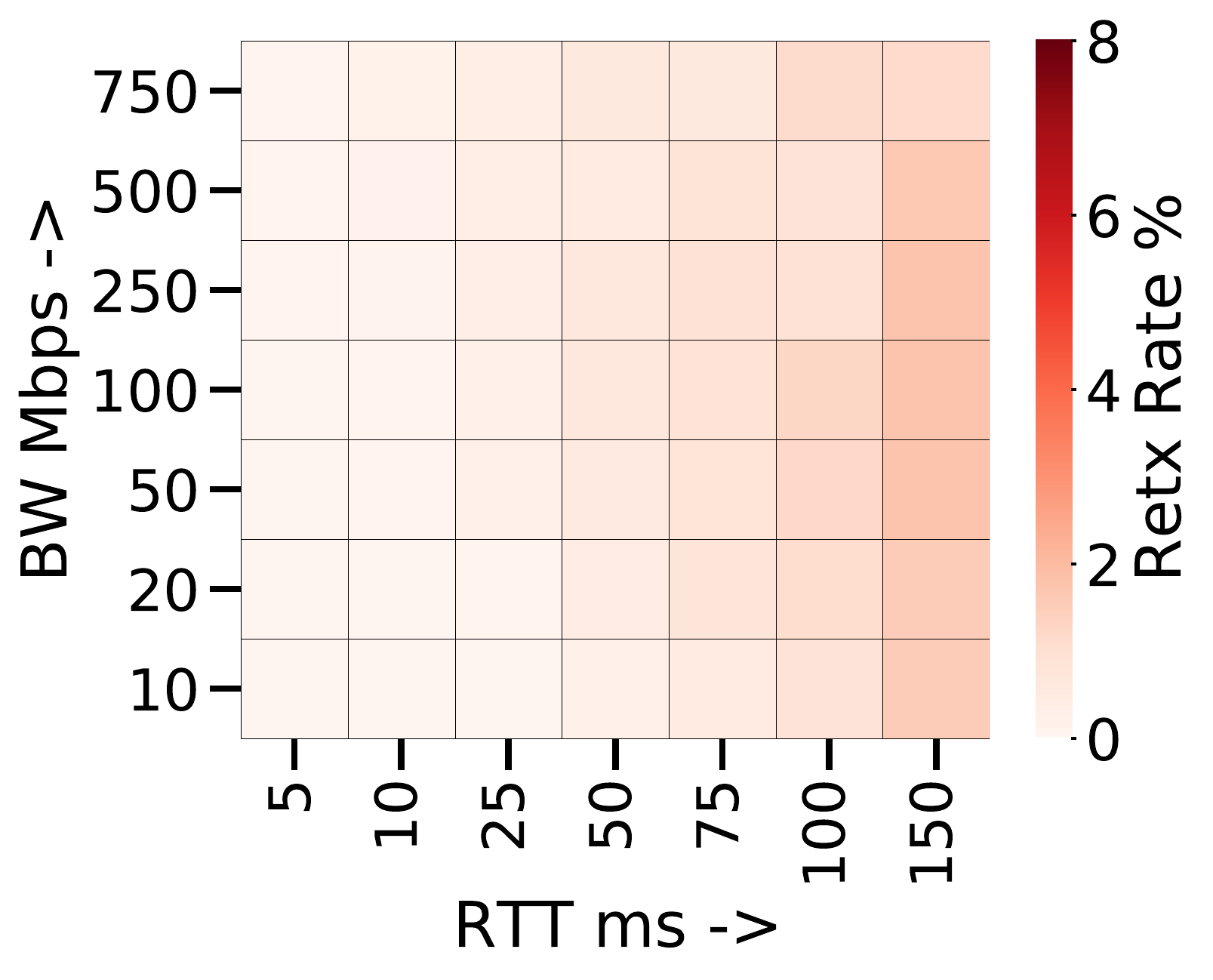}
            \subcaption{BBRv2}
            \label{fig:retx_bbr2}
        \end{minipage}
        \caption{The heatmap of the retransmission rate of BBR/BBRv2 under various network conditions. The numbers in squares are retransmission rates in percentage.}
        \label{fig:retx_bbr_vs_bbr2}
    \end{minipage}
\end{figure}

\subsection{Retransmission and throughput}
\label{sec:retx_vs_tput}


As one of BBRv2's design goals is to reduce unnecessary retransmissions in shallow-buffered networks, next we investigate whether BBRv2 achieves this design goal. The experimental setup is similar to that in the previous work~\cite{cao_when_2019}, where the bottleneck bandwidth varies in 10$\sim$750~Mbps and the path RTT varies in 5$\sim$150~ms as these values are commonly employed in modern networks~\cite{cao_when_2019, hock_experimental_2017, Huffaker2002DistanceMI}. The buffer size at the bottleneck link is set to 100KB to emulate a shallow-buffered network because 100KB is less than the BDP of most bandwidth-RTT combinations in our setup. One TCP flow from H1 to H3 runs for 30 seconds and the retransmission rate is recorded for each setup. 

\begin{figure}
    \begin{minipage}{1\linewidth}
        \begin{minipage}[t]{0.48\linewidth}
            \centering
            \includegraphics[width=1\linewidth]{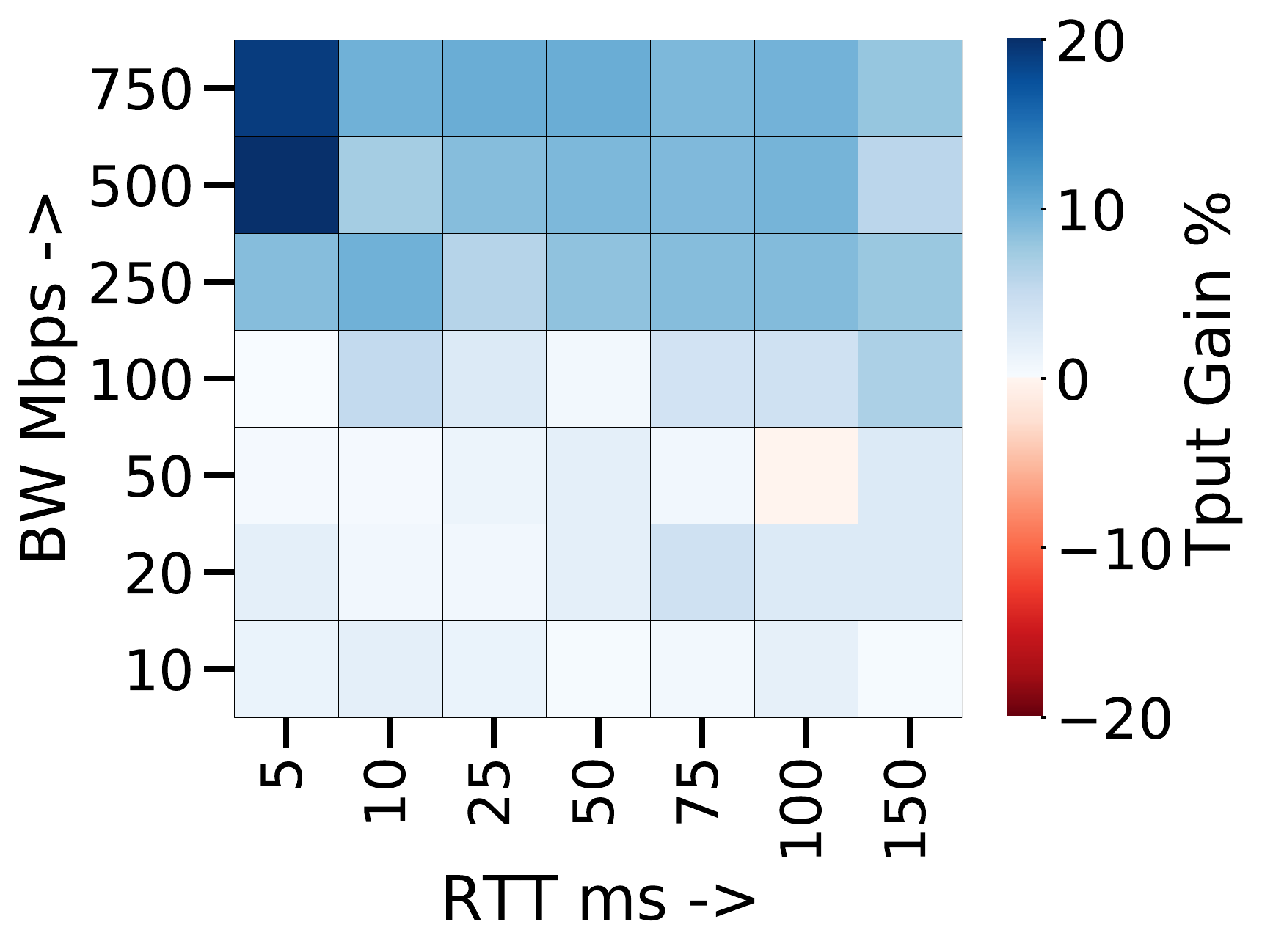}
            \subcaption{100KB buffer}
            \label{fig:tput_bbr_vs_bbr2_shallow}
        \end{minipage}
        \hfill
        \begin{minipage}[t]{0.48\linewidth}
            \centering
            \includegraphics[width=1\linewidth]{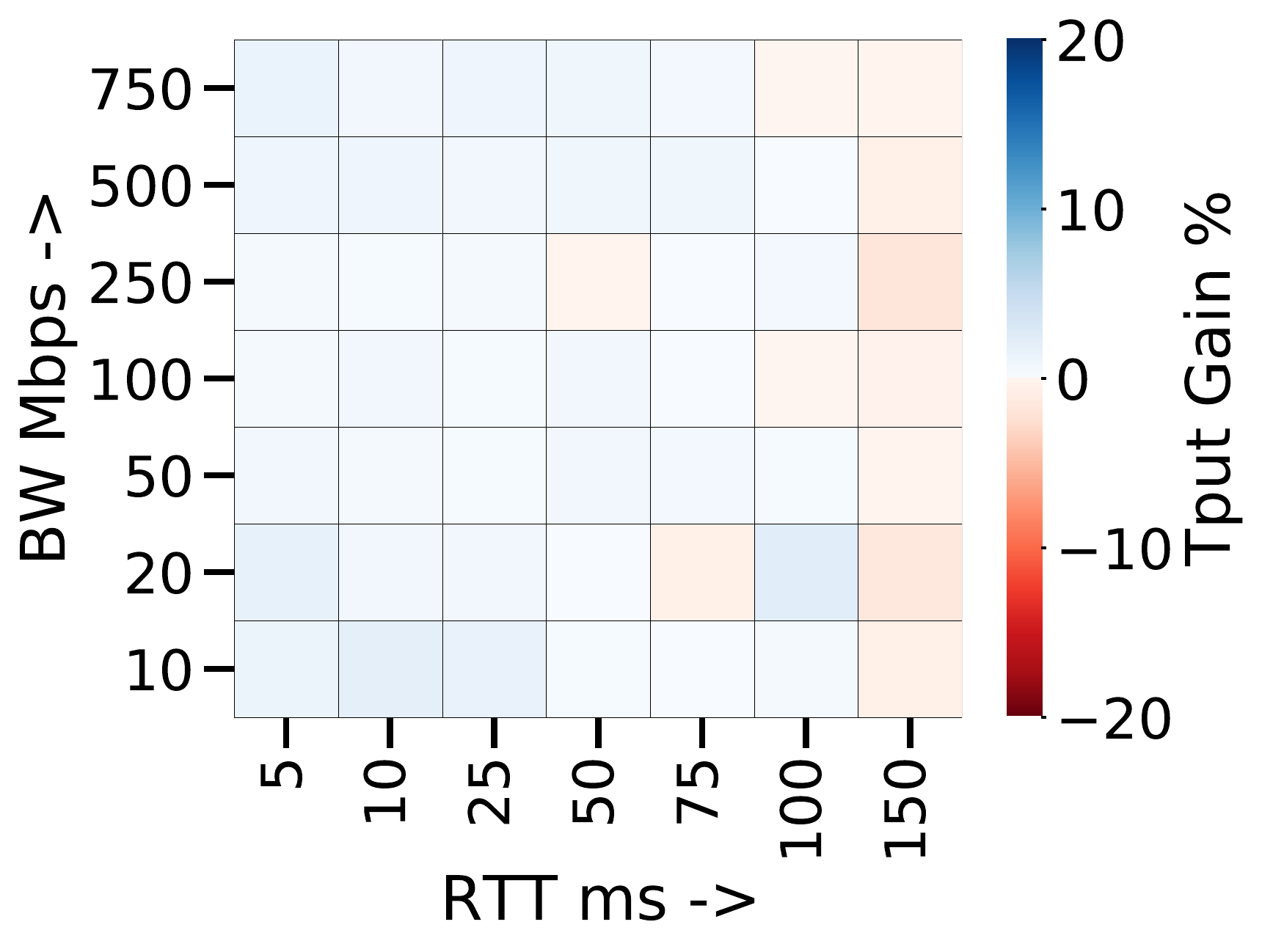}
            \subcaption{10MB buffer}
            \label{fig:tput_bbr_vs_bbr2_deep}
        \end{minipage}
        \caption{The heatmap of \textit{Tput\_Gain} (in percentage) in (a) shallow and (b) deep buffered networks.}
        \label{fig:tput_bbr_vs_bbr2}
    \end{minipage}
\end{figure}

The heatmaps in Fig.~\ref{fig:retx_bbr_vs_bbr2} show the retransmission rates of BBR and BBRv2 under various network conditions. We can observe that the retransmission rate of BBRv2 is significantly reduced compared with that of BBR, especially when the BDP is larger than 400KB (i.e. the buffer size $\leq$ 0.25$\times$ BDP).

The lower retransmission rate of BBRv2 in shallow-buffered networks stems from the fact that BBRv2 reacts to packet losses, while BBR does not. In the Startup and ProbeUp state, BBRv2 tries to send at a rate higher than the bottleneck bandwidth, which leads to excessive losses (i.e. loss rate $\geq$ 2\%). The excessive losses trigger \emph{inflight\_hi} to be set to the current inflight size that is likely close to BDP in shallow-buffered networks. As a result, BBRv2's inflight size is bounded beblow 0.85$\times$ \emph{inflight\_hi} in ProbeCruise because it tries to leave headroom for other flows to explore bandwidth. Since BBRv2 flows spend most of their lifecycle in ProbeCruise, the average throughput of BBRv2 is expected to be 15\% lower than the available bandwidth.

That said, BBRv2 trades off throughput against retransmission in shallow-buffered networks. To verify this, we compute the throughput gain of BBR over BBRv2 (\emph{Tput\_Gain}), which is defined in Eq.~\ref{eq:tput_gain}, where $Tput_{BBR}$ (resp. $Tput_{BBRv2}$) is the average throughput of a BBR (resp. BBRv2) flow over 30 seconds. 

\begin{equation}
    \centering
    Tput\_Gain = \frac{Tput_{BBR} - Tput_{BBRv2}}{Tput_{BBRv2}}
    \label{eq:tput_gain}
\end{equation}

Fig.~\ref{fig:tput_bbr_vs_bbr2_shallow} plots the \emph{Tput\_Gain} under various network conditions. We can observe that in the network conditions where BBRv2 reduces the retransmission rate (when the BDP exceeds 400KB), BBRv2 achieves lower throughput than BBR. Specifically, the throughput of BBRv2 is 13\%$\sim$16\% lower than that of BBR in these cases, which coincides with our analysis.
 
 

In deep-buffered networks, however, the packet losses are much less often. It is thus expected that the throughput of BBR and BBRv2 are comparable. This is confirmed by the results in Fig.~\ref{fig:tput_bbr_vs_bbr2_deep}, where the buffer size is configured at 10MB, larger than the BDP of most of the bandwidth-RTT combinations in our setup. The throughput differences between BBR and BBRv2 are indeed marginal in these networks.

\subsection{Resilience to random losses}
\label{sec:measurement:loss_resilience}

Several early tests~\cite{cardwell_bbr_105, kfoury_emulation-based_2020, song_understanding_2021} have shown that BBRv2 is less resilient to random losses than BBR, since BBRv2 limits its inflight size by the \emph{inflight\_lo} and \emph{inflight\_hi}, which both react to all types of losses. In BBRv2, there are two parameters that decide how the \emph{inflight\_lo} and \emph{inflight\_hi} react to losses. One is the explicit loss threshold ($\alpha$) and the other one is the \emph{inflight\_lo} reduction factor ($\beta$). In our experiments, we investigate BBRv2 variants with different $\alpha$ and $\beta$ under random loss, where each specific BBRv2 variant is referred as BBRv2($\alpha$, $\beta$). For $\alpha$, we cap it at 20\% to match the maximum loss rate that BBR can tolerate; for $\beta$, we only evaluate the difference between the case with (i.e. $\beta=0.3$) and without it i.e. ($\beta=0$)\footnote{The default value 0.3 is necessary for BBRv2 to co-exist with Cubic~\cite{bbr2_kernel}}.

In the experiment, the bottleneck bandwidth is set to 40Mbps, and the path RTT is 40ms. The buffer size is set to 32$\times$ BDP to avoid packet loss due to buffer overflow. The random loss rate ranges from 0\% to 30\%.

\begin{figure}
    \centering
    \includegraphics[width=0.85\linewidth]{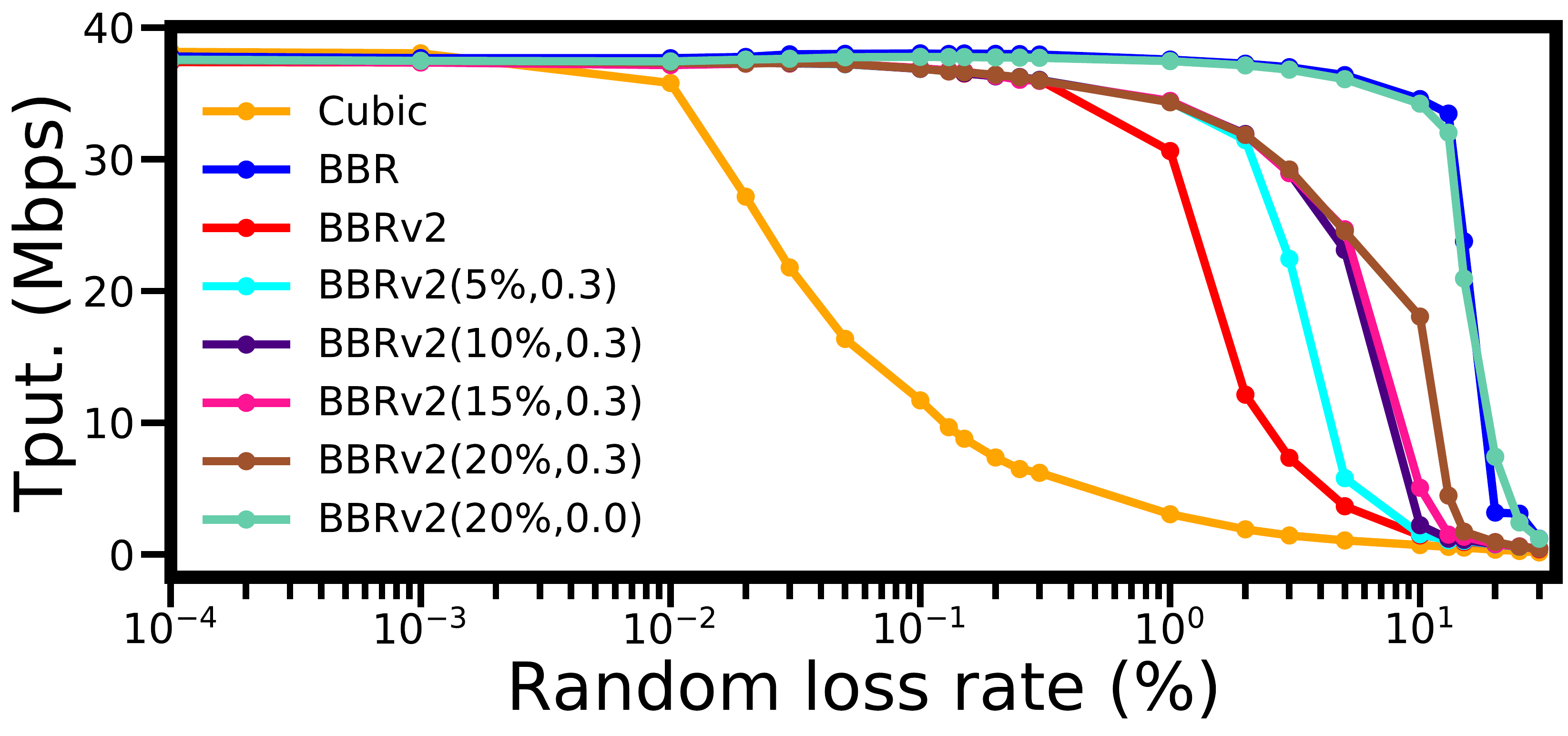}
    \caption{Avg. throughput against different random loss rates (buffer size = 32$\times$ BDP).}
    \label{fig:loss_resilience_measurement}
\end{figure}

Fig.~\ref{fig:loss_resilience_measurement} reports the average throughput of each CCA against random loss rates. We observe that the throughput of BBRv2 drops significantly after the random loss rate reaches 2\%. There is a clear sign that the $\alpha$ impacts the loss resilience of BBRv2: as the $\alpha$ increases, we can observe the improvement of loss resilience of BBRv2. For instance, with a 10\% random loss rate, BBRv2(20\%, 0.3) reaches around half of the maximum bandwidth while BBRv2's throughput nearly drops to zero. The impact of $\beta$ is also remarkable: the loss resilience of BBRv2(20\%, 0.3) is lower than that of BBRv2(20\%, 0) that performs similar to BBR. 


The above results indicate that BBRv2's loss resilience can be improved via raising the $\alpha$. Yet, there is a concern --- how does the $\alpha$ impact the retransmission rate in shallow-buffered networks as we already saw that BBRv2 alleviates the retransmission issue by setting \emph{inflight\_hi} upon the loss rate exceeding $\alpha$ to lower down its inflight size (see \S\ref{sec:retx_vs_tput}).






\begin{figure}
    \centering
    \includegraphics[width=0.85\linewidth]{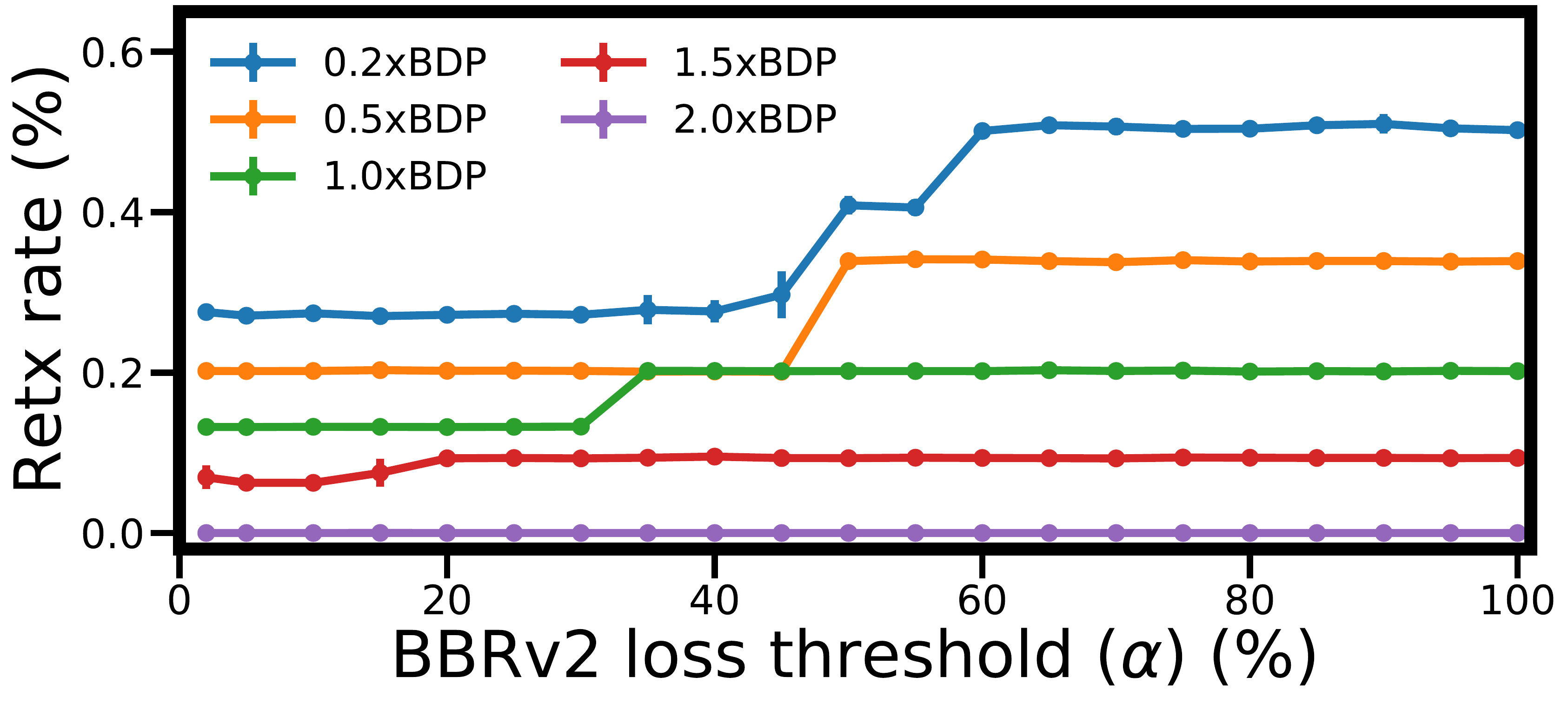}
    \caption{Retransmission rates versus loss thresholds ($\alpha$) under networks with various buffer sizes. The errorbars in the figure represent the standard deviations of retransmission rate. Note that the $\beta$ was fixed at 0.3 for all experiments; the default $\alpha$ in BBRv2 is 2\% (the first data point of every line).}
    \label{fig:retx_vs_alpha}
\end{figure}

To investigate the aforementioned concern, we further extended the experiments by considering more BBRv2 variants ($\alpha$ $\in$ [2\%, 100\%], $\beta$ = 0.3) and more configurations on buffer size (buffer size $\in$ \{0.2, 0.5, 1.0, 1.5, 2.0\}$\times$ BDP). Fig.~\ref{fig:retx_vs_alpha} plots the retransmission rates of all those BBRv2 variants under 0\% random loss rate (to eliminate the impact of random losses on retransmission rate), which shows the impact of buffer size. Two observations are notable. Firstly, we observe that the retransmission rate increases when the $\alpha$ exceeds a certain point, which depends on the bottleneck buffer size. The $\alpha$ values beyond the turning points are too high to be reached by the temporary loss rate, thus, limiting the efficacy of \emph{inflight\_hi}. Secondly, if the buffer size is large enough (i.e. 2$\times$ BDP in our experiments), the retransmissions are eliminated, thus, the value of $\alpha$ becomes irrelevant. 

\begin{figure}
    \centering
    \includegraphics[width=0.6\linewidth]{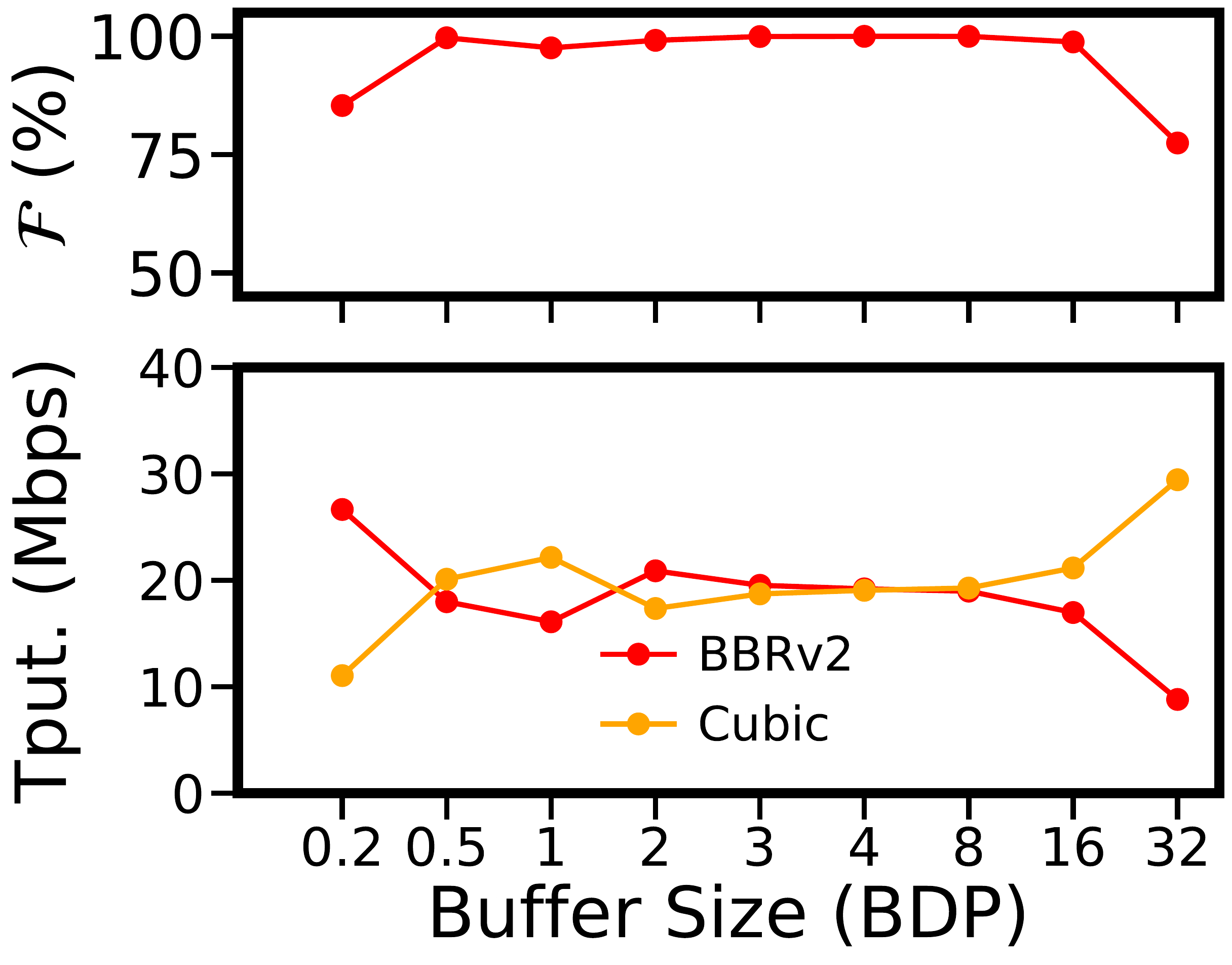}
    \caption{Inter-protocol fairness of BBRv2(20\%, 0.3).}
    \label{fig:cc_fair_bbr2_20alpha}
\end{figure}

Another concern about lifting $\alpha$ is the impact on the inter-protocol fairness because the larger $\alpha$ is, the slower reaction of BBRv2 to losses is, which makes BBRv2 more aggressive to loss-based CCAs. To investigate this concern, we test the inter-protocol fairness of BBRv2(20\%, 0.3) using the same setup in \S\ref{sec:measurement:inter_fair}, and plot the results in  Fig.~\ref{fig:cc_fair_bbr2_20alpha}. In comparison with Fig.~\ref{fig:cc_fairness_bbr2}, we can see that the inter-protocol fairness of BBRv2 is indeed worsened in the case of extremely shallow buffer (0.2$\times$ BDP) due to the increased aggressiveness caused by a larger $\alpha$. Nevertheless, we also observe that the fairness index is improved under moderate buffers because the increased aggressiveness also makes BBRv2 less vulnerable to Cubic when the bottleneck buffer becomes larger.


\vspace{0.5em}
\noindent\textbf{Summary of random loss resilience:} The loss resilience of BBRv2 can be improved by raising the loss threshold $\alpha$. Nevertheless, the threshold $\alpha$ should be carefully tuned according to the bottleneck buffer size to avoid increasing retransmissions and being too aggressive to loss-based CCAs in extremely shallow-buffered networks. 



\subsection{Responsiveness to bandwidth dynamics}
\label{sec:measurement:bw_dynamics}


In networks with highly dynamic available bandwidth~\cite{wang_active-passive_2019, winstein_stochastic_nodate, li_measurement_2018}, BBRv2's bandwidth probing may fail to quickly adapt to bandwidth changes. Next, we investigate BBRv2's responsiveness to bandwidth changes.

\begin{figure*}[thb]
        \centering
        \begin{minipage}[t]{0.45\linewidth}
            \centering
            \includegraphics[width=1\linewidth]{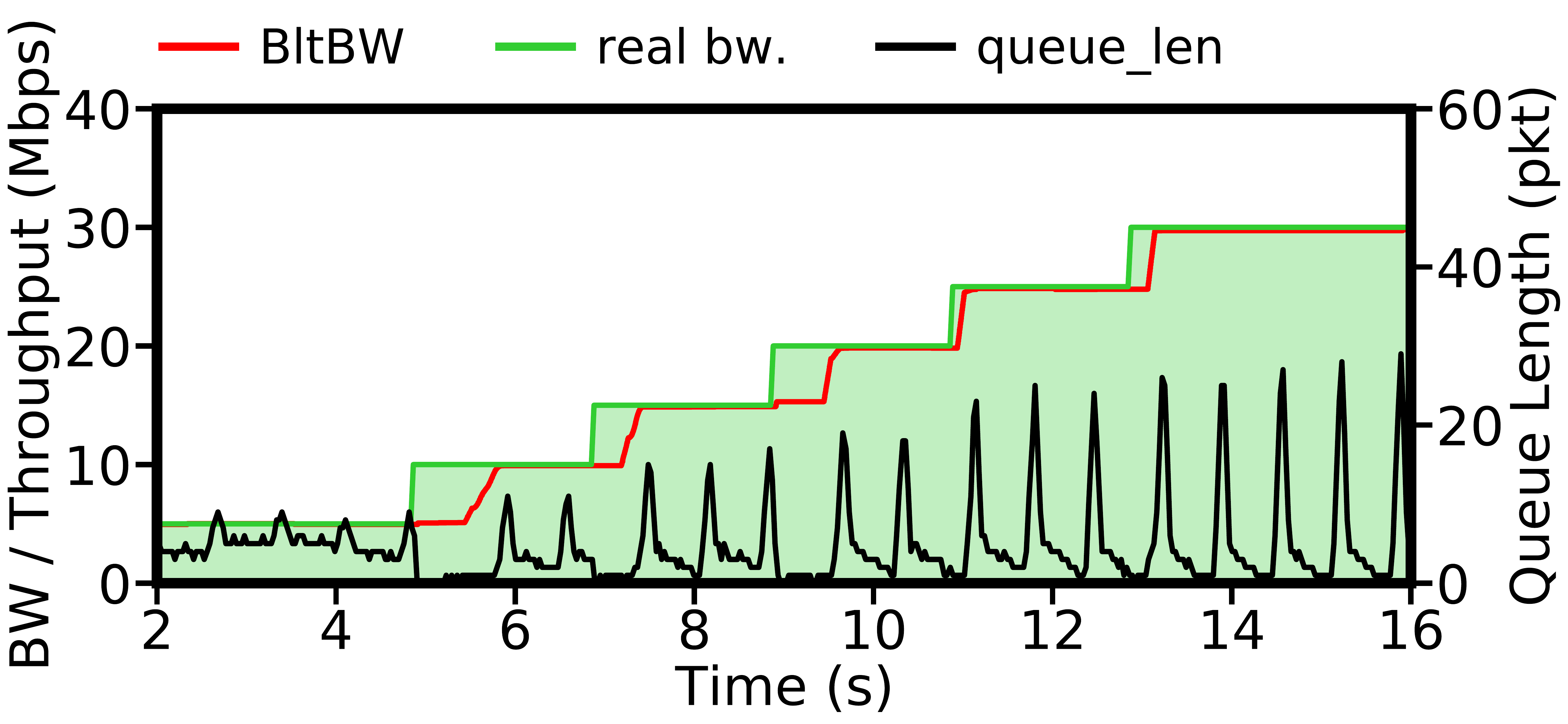}
            \subcaption{BBR (bw. increasing)}
            \label{fig:bw_inc_bbr}
        \end{minipage}
        \hfill
        \begin{minipage}[t]{0.45\linewidth}
            \centering
            \includegraphics[width=1\linewidth]{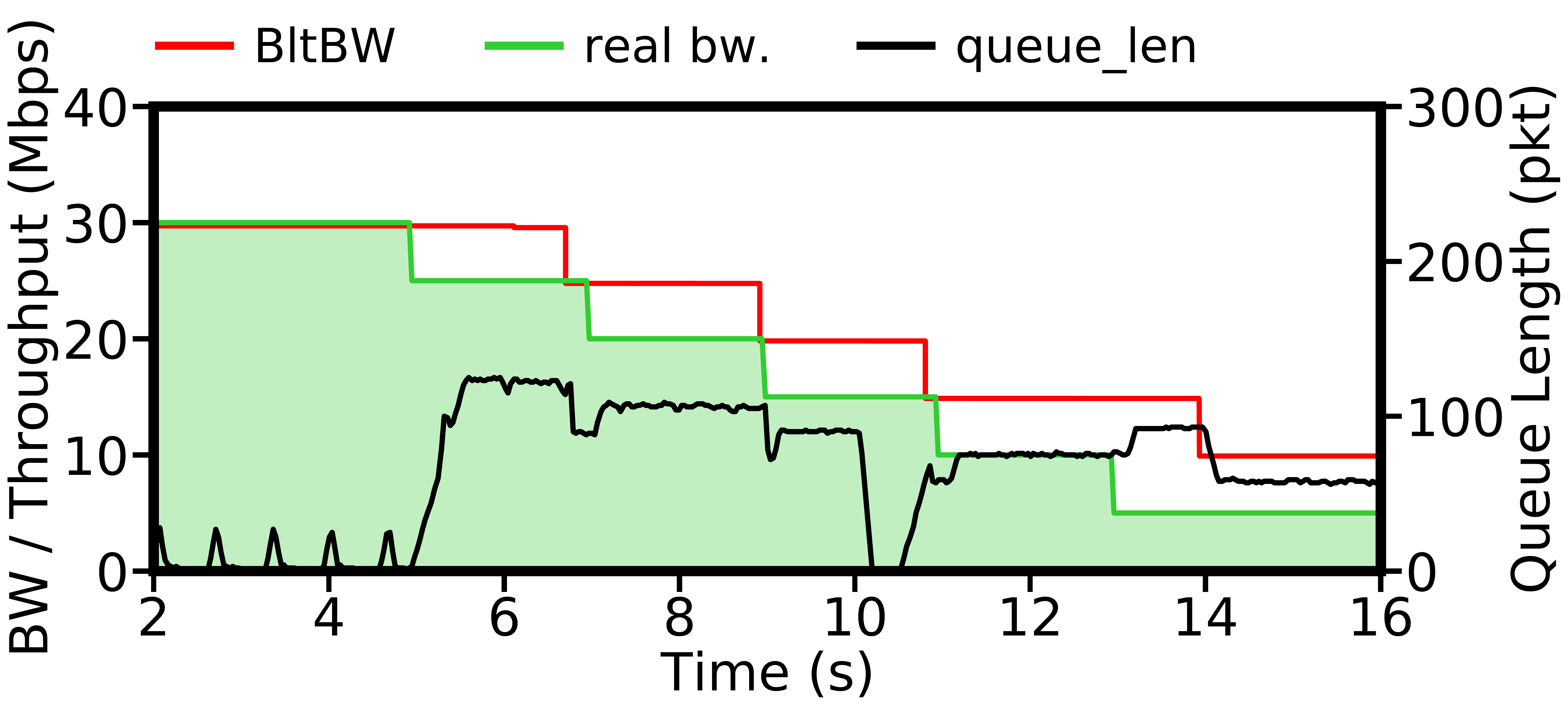}
            \subcaption{BBR (bw. decreasing)}
            \label{fig:bw_dec_bbr}
        \end{minipage}
        \hfill
        \begin{minipage}[t]{0.45\linewidth}
            \centering
            \includegraphics[width=1\linewidth]{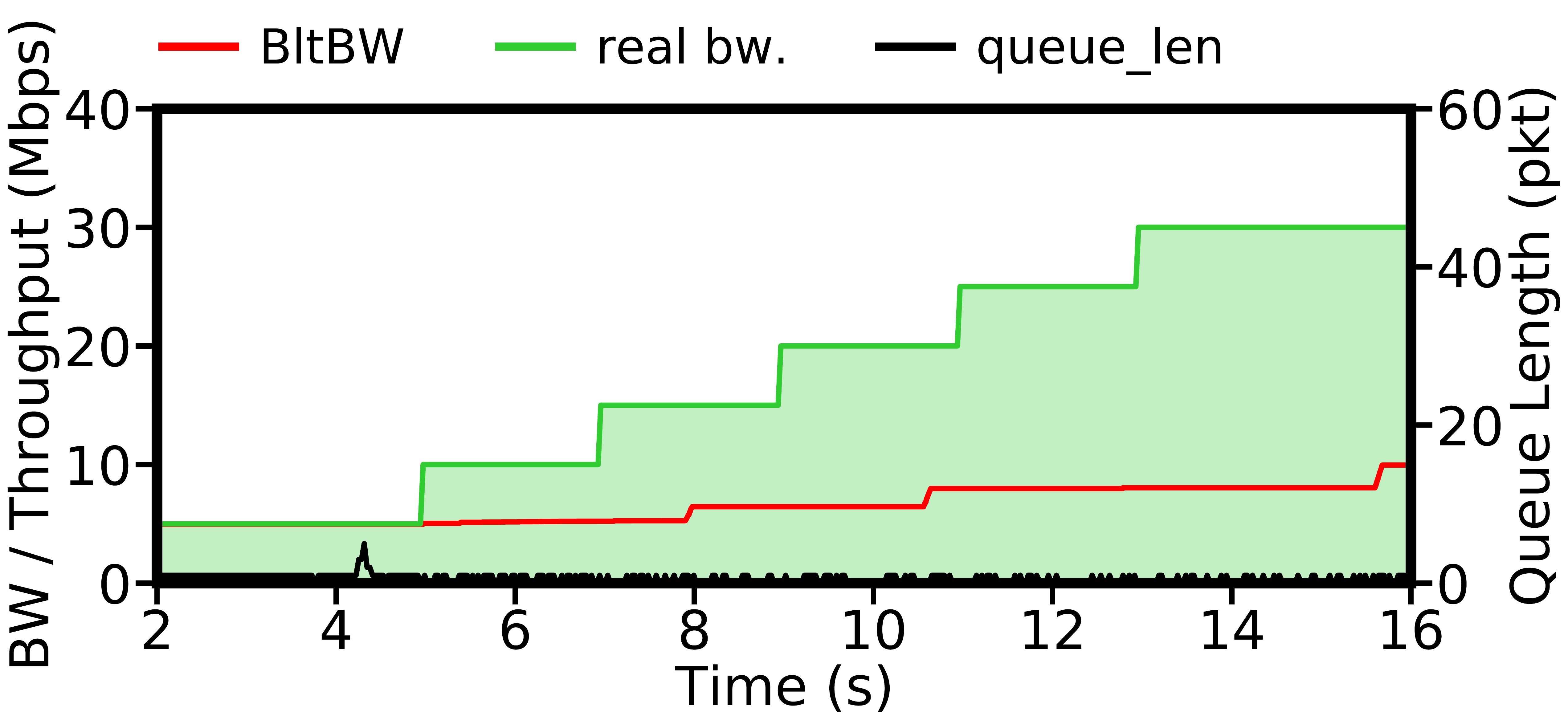}
            \subcaption{BBRv2 (bw. increasing)}
            \label{fig:bw_inc_bbr2}
        \end{minipage}
        \label{fig:bw_inc}
        \hfill
        \begin{minipage}[t]{0.45\linewidth}
            \centering
            \includegraphics[width=1\linewidth]{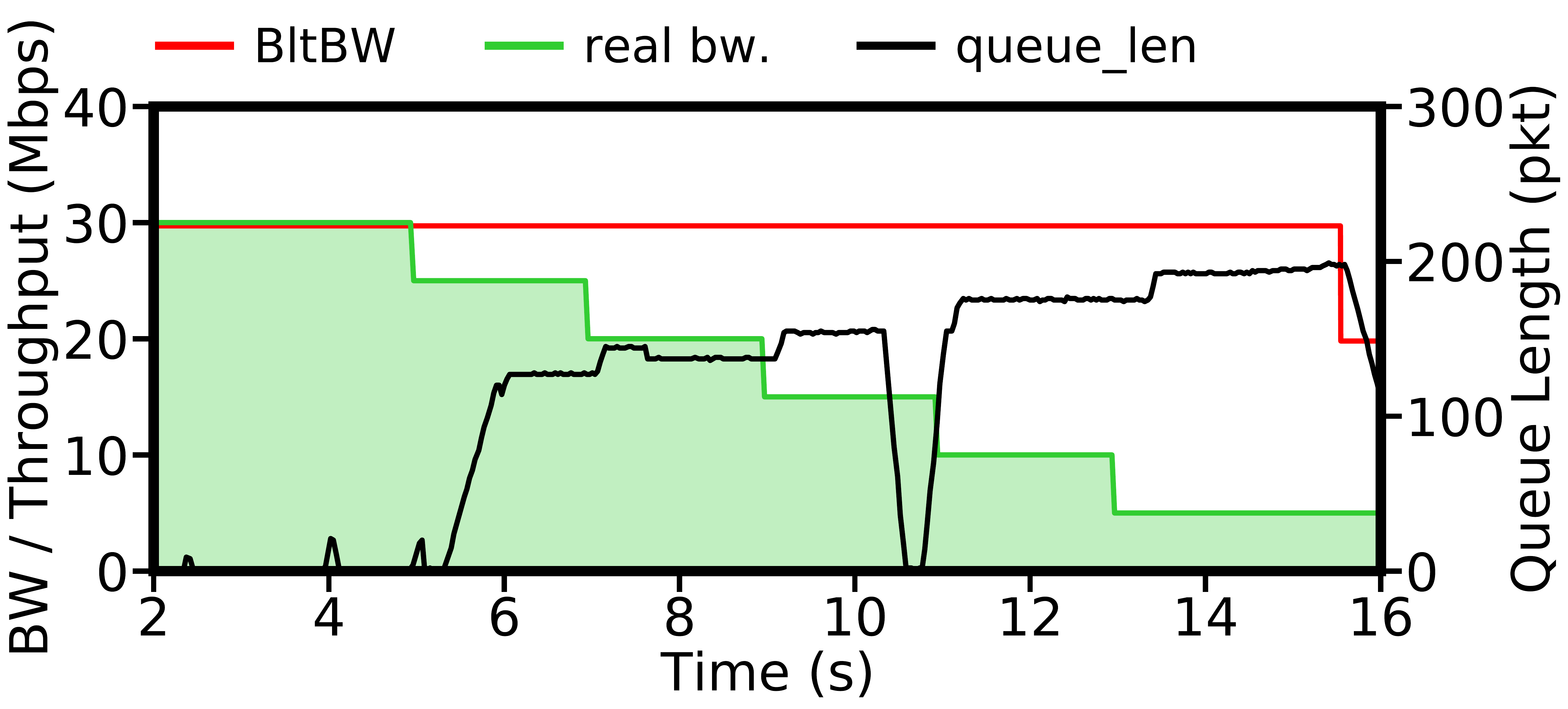}
            \subcaption{BBRv2 (bw. decreasing)}
            \label{fig:bw_dec_bbr2}
        \end{minipage}
        \caption{Responsiveness to bandwidth increases (a, c) and decreases (b, d). The red line represents the \textit{BtlBW} estimation of BBR/BBRv2, and the green line indicates the real bandwidth of the bottleneck. The dark line shows the dynamics of the bottleneck link's queue length. Note that the spikes of queue length in (a) and (c) are caused by the periodical bandwidth probing of BBR/BBRv2, and the sudden drop of queue\_len around 10s in (b) and (d) is because BBR/BBRv2 enters the ProbeRTT state.}
        \label{fig:resp_to_bw_dynamics}
\end{figure*}


The experiments are designed as follows. The bandwidth of the bottleneck link is configured to increase or decrease 5Mbps every 2 seconds, the path delay is set to 40ms, and the buffer size is set to 32$\times$ BDP. The internal variables during flow transmission (including \textit{pacing\_rate} and \textit{BtlBW}, instantaneous throughput, and the queue length at the bottleneck link) are sampled at an interval of 100ms.

Fig.~\ref{fig:resp_to_bw_dynamics} shows how BBR and BBRv2 adapt to bandwidth increases or decreases respectively. In this figure, the upward and downward spikes of queue length correspond to the actions of BBR/BBRv2 in probing for more bandwidth or draining the bottleneck buffer. We can observe that BBRv2 is less effective than BBR in terms of responsiveness to bandwidth dynamics, resulting in low utilization of bandwidth and long queuing delay.


As we discussed in \S\ref{sec:background}, to match the interval between Reno loss recovery epochs for better inter-protocol fairness, BBRv2 uses $min\{rand(2,3),\frac{BDP}{MSS}\times{}RTT\}$ seconds as its probing interval. This interval can be tens of RTTs, which is too conservative in such a dynamic environment. That said, BBRv2 improves the inter-protocol fairness, at the cost of poorer responsiveness to bandwidth dynamics.

\subsection{Resilience to network jitters}
\label{sec:measurement:jitter}

Several works~\cite{wang_active-passive_2019, kumar_tcp_2019} have shown that throughput collapse occurs when BBR operates in high-jitter networks that are widely deployed, e.g. WiFi and 5G networks operating in mmWave band~\cite{zhang_will_2018, chitimalla_5g_2017, kumar_tcp_2019}, and cellular networks~\cite{kumar_tcp_2019,  beshay_link-coupled_2017} especially when high-mobility involves such as high-speed rails~\cite{wang_active-passive_2019}. It is interesting to investigate whether BBRv2 operates well in networks with high jitters.

In this experiment, the bottleneck bandwidth is 40Mbps and the path RTT is 40ms. The bottleneck buffer size is set to 32$\times$ BDP to avoid buffer overflow. To emulate jitters, \emph{tc} is used to add jitters following Gaussian distribution at R3's interface that connects to H3. The mean value of the Gaussian distribution varies from 0$\sim$120~ms to emulate different degrees of jitters.

\begin{figure}
    \centering
    \begin{minipage}{0.6\linewidth}
        \centering
        \includegraphics[width=1\linewidth]{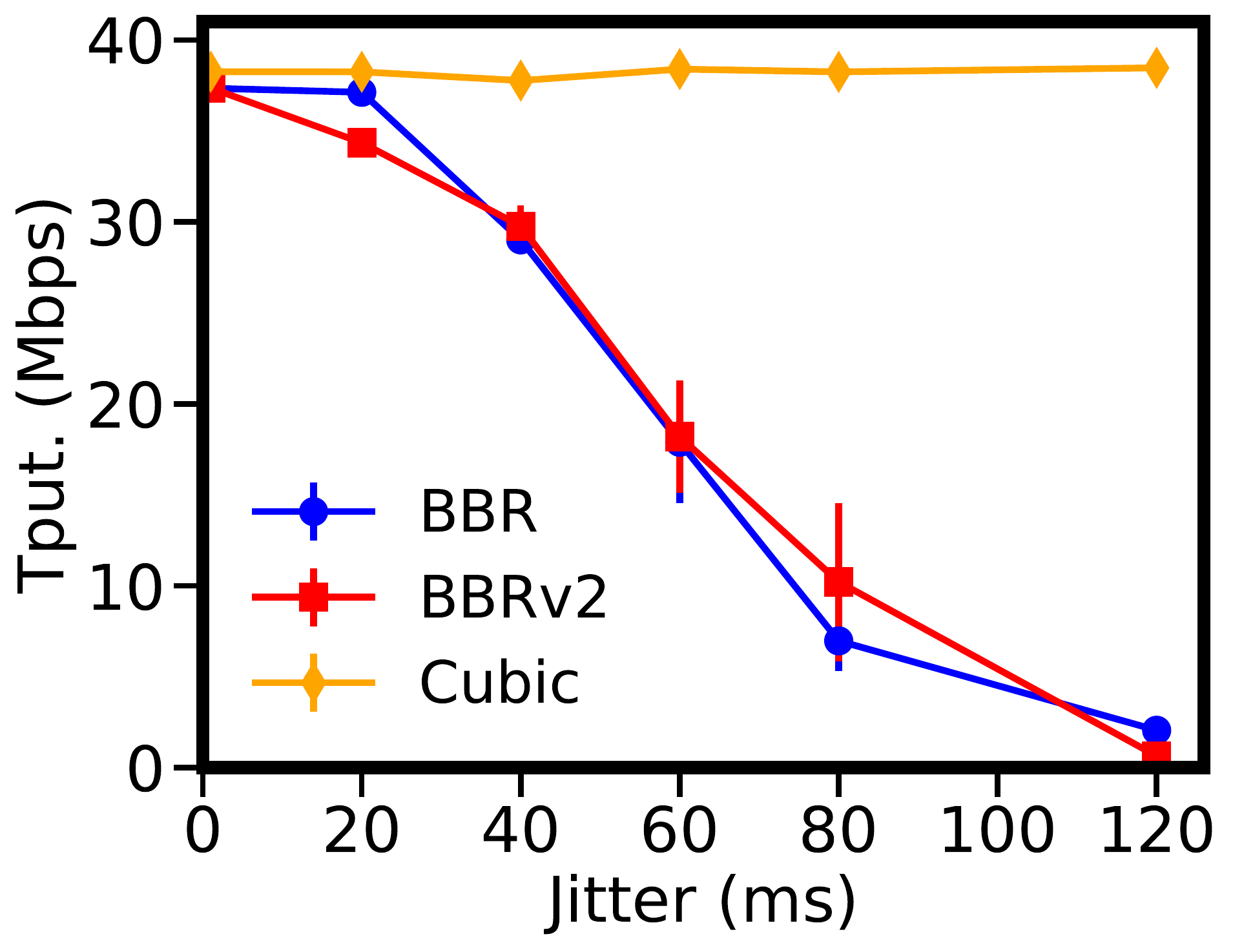}
    \end{minipage}
    \caption{Avg. throughput against different levels of jitters. $x=0$ is equivalent to no jitters.}
    \label{fig:measure_resilience_to_jitters}
\end{figure}

Fig.~\ref{fig:measure_resilience_to_jitters} shows the average value and the standard deviation of throughput of Cubic, BBR, and BBRv2 under various levels of jitter. Compared with Cubic, both BBR and BBRv2 experience low throughput under high jitters. As documented by Kumar et al.~\cite{kumar_tcp_2019}, BBR underestimates \textit{RTprop} in such networks because it uses a recent 10s minimum RTT to approximate \textit{RTprop}, leading to \textit{cwnd} exhaustion. This problem still exists in BBRv2, even if BBRv2 updates \textit{RTprop} 2$\times$ frequently than BBR (i.e. BBRv2 uses the minimum RTT in recent 5s to estimate \textit{RTprop}). We also note that the significant throughput degradation starts when the average jitter reaches the path RTT without jitter (i.e. 40ms).

\subsection{Summary and Implication}

We observe that BBRv2 improves the inter-protocol fairness and RTT fairness, and reduces retransmission rates under shallow buffers, at the cost of slow responsiveness to bandwidth dynamics and low resilience to random loss. 

First, the root cause for the slow responsiveness is that BBRv2 is over-conservative regarding bandwidth probing. That said, it fails to achieve a good balance between the aggressiveness in probing for more bandwidth and the fairness against loss-based CCAs. Note that, however, recklessly increasing BBRv2's aggressiveness in bandwidth probing may lead BBRv2 to generate overwhelming retransmissions and unfairly share bandwidth with loss-based CCAs. In the next section, we propose BBRv2+, which incorporates \signame{} to cautiously guide the aggressiveness of bandwidth probing to avoid reducing the fairness against loss-based CCAs. The challenge is how to effectively use this signal and how to avoid being suppressed by other loss-based CCAs in deep-buffered networks as other delay-based CCAs. 

Second, the resilience to random loss can be improved by raising the loss threshold $\alpha$, where the value of $\alpha$ needs to be set according to the bottleneck buffer size.

Last, the throughput degradation of BBR and BBRv2 in high-jitter networks is own to the underestimation of \textit{RTprop}, which in turn leads to a smaller estimation of BDP. We propose a compensation mechanism of BDP that enables the estimated BDP to be close to the real BDP.

%% file: sections/motivation-arch.tex


\label{sec:design}

\begin{table*}[ht]
    \centering
    \begin{tabularx}{\linewidth}{l|X}
        \toprule
        Variable & Functionality \\
        \midrule
         $\mathrm{MinRTT_{prev\_rtt}}$ & The minimum RTT measured in the previous RTT round.  \\
         $\mathrm{MinRTT_{curr\_rtt}}$ & The minimum RTT measured in the current RTT round. \\
         $\mathrm{MinRTT_{before\_probe}}$ & Saving the $\mathrm{MinRTT_{curr\_rtt}}$ before entering ProbeUp.  \\
         $\mathrm{MinRTT_{curr\_cruise}}$ & The minimum RTT measured in the current ProbeCruise state. \\

         \multirow{2}{*}{$\mathrm{Max_{4RTT}(jitter)}$} & The \textit{max\_filter} tracking the maximum jitter in recent four RTT rounds. Here, the jitter is equivalent to the RTT variation maintained by TCP. \\

         \bottomrule
    \end{tabularx}
    \caption{The new state variables in BBRv2+}
    \label{tab:bbrv2plus_vars}
\end{table*}

\begin{figure}
    \centering
    \includegraphics[width=\linewidth]{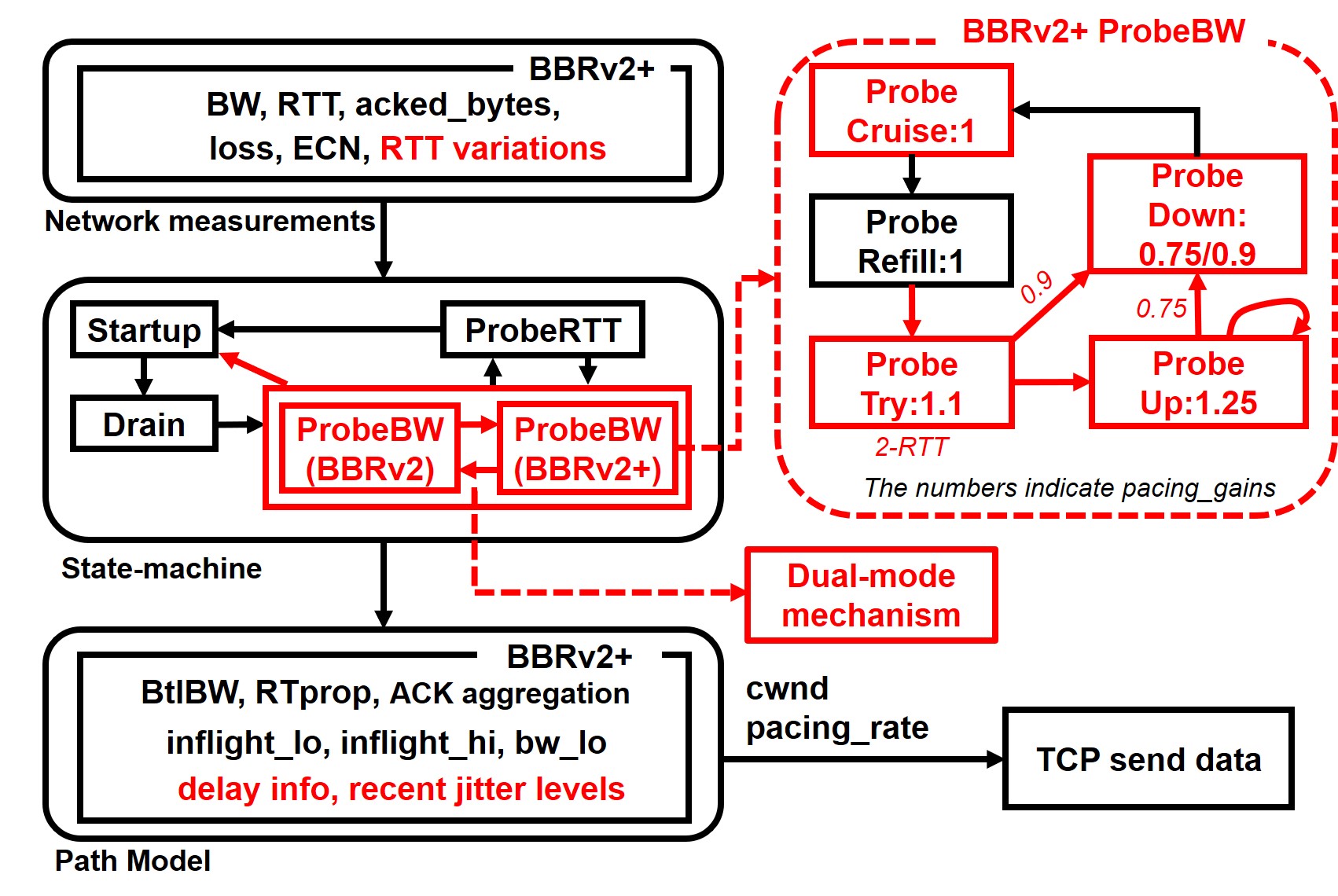}
    \caption{BBRv2+ architecture. The parts that differs from BBRv2 are highlighted in red color.}
    \label{fig:bbrv2plus_arch}
\end{figure}




Motivated by our measurement results, we design and implement BBRv2+, in order to address the pitfalls of BBRv2 while maintaining its advantages over BBR (i.e. improved fairness and reduced retransmissions in shallow-buffered networks). The basic idea is to incorporate \signame{} in BBRv2+'s path model to balance between the aggressiveness in probing for more bandwidth and the fairness against loss-based CCAs (\S\ref{sec:design:probebw}). That said, BBRv2+ tries to be more aggressive than BBRv2, where the aggressiveness is guided by the \signame{}. As the use of the \signame{} may lead BBRv2+ to perform poorly when it co-exists with loss-based CCAs, a dual-mode mechanism is introduced in BBRv2+, where BBRv2+ switches to use BBRv2's state-machine (i.e. invalidating the effect of the \signame{}) or returns back to use the redesigned state-machine depending on whether loss-based competitors co-exist (\S\ref{sec:design:coexist}). Moreover, BBRv2+ compensates the estimated BDP when detecting high jitters in order to get an accurate estimation of BDP. (\S\ref{sec:design:comp_cwnd}).





\subsection{Overview}

The architecture of BBRv2+ is shown in Fig.~\ref{fig:bbrv2plus_arch}. BBRv2+ incorporates \signame{} in its path model. Specifically, the \signame{} consists of three state variables (the first three variables listed in Table.~\ref{tab:bbrv2plus_vars}) of minimum RTTs, which reflect the change of queuing delay over time. 
The \signame{} facilitates quick responsiveness to bandwidth dynamics. Particularly, a new sub-state, ProbeTry, is added into the ProbeBW state. In ProbeTry, BBRv2+ slightly speeds up to examine if this acceleration will lead to increased RTTs. In the case of increased RTTs, BBRv2+ quits this probing and moves to the ProbeDown state to drain the queue at the bottleneck link; otherwise, it moves to the ProbeUp state to further explore available bandwidth. BBRv2+ also uses the \signame{} to quickly adapt to bandwidth decreases---it quickly updates its bottleneck bandwidth estimation to the current bandwidth measurement if an obvious increase of RTT is observed when BBRv2+ is not probing for bandwidth.

Like other CCAs that use delay-based signals, BBRv2+ will be suppressed when co-existing with loss-based CCAs under deep buffers~\cite{al-saadi_survey_2019}, as the loss-based CCAs constantly fill the buffer, leading BBRv2+ to falsely yield up obtained bandwidth. BBRv2+ uses a dual-mode mechanism that forces BBRv2+ to use BBRv2's state-machine when loss-based CCAs co-exist.

Finally, BBRv2+ uses a BDP compensation mechanism to address the \textit{cwnd} exhaustion problem caused by network jitters. Our key observation is that in high-jitter networks, the BDP will be underestimated because of the underestimation of \textit{RTprop}. The mechanism compensates BDP by taking the recent RTT variations into consideration; this compensation mitigates the underestimation issue significantly.

\subsection{Redesign of the ProbeBW state}
\label{sec:design:probebw}

In the case of bandwidth increments, BBRv2+ needs to start probing for more bandwidth quickly instead of spending time on cruising with the current estimated \textit{BtlBW}. Thus, the probing interval needs to be reasonably shortened, which is set to approximately match the probing interval of BBR (8 rounds of RTT). However, if BBRv2+ is already sending at the speed close to bottleneck bandwidth, the probing interval above may result in more packet losses and thus unfairness against loss-based CCAs in shallow-buffered networks.

Thus, a two-step probing mechanism incorporating the \signame{} is introduced in BBRv2+, as shown in Fig.~\ref{fig:bbrv2plus_arch}. A new sub-state ProbeTry, which lasts for two RTTs, is inserted before entering ProbeUp in the state machine. In the first RTT of ProbeTry, BBRv2+ slightly increases its \textit{pacing\_rate} by increasing \emph{pacing\_gain} to 1.1. In the second RTT, BBRv2+ reduces \emph{pacing\_gain} to 1.0 and monitors if $\mathrm{MinRTT_{curr\_rtt}}$ is larger than $\gamma\times\mathrm{MinRTT_{prev\_rtt}}$, where MinRTT is measured on the ACKs for the packets sent in the previous round, thus, reflecting the queuing delay caused by the previous round (see Table~\ref{tab:bbrv2plus_vars}). The rationale of using $\gamma>1$ is to introduce a relaxing factor tolerate noises in RTT measurements, where a small $\gamma$ may lead BBRv2+ to miss some chances to explore bandwidth while a large $\gamma$ may make BBRv2+ over-aggressive. In our current implementation, we set $\gamma=1.02$ to tolerate noises for 2\% of RTT measurements. It is worth noting that $\gamma$ is a design parameter and can be tuned by designers\footnote{All the design parameters of BBRv2+ in our current implementation are exposed to user-space through the \texttt{/sys/module} interfaces, enabling designers to change the parameters without recompiling the kernel module.}. If $\mathrm{MinRTT_{curr\_rtt}}$ $>$ $\gamma \times \mathrm{MinRTT_{prev\_rtt}}$, BBRv2+ transits to ProbeDown with \emph{pacing\_gain} as 0.9 to drain the queue accumulated during the first RTT of ProbeTry. Otherwise, it enters ProbeUp to probe for more bandwidth. 

To further boost the speed of bandwidth discovery, BBRv2+ also incorporates a continuous probing mechanism based on the \signame{} \fy{Mark}. Specifically, at the end of ProbeUp, if the $\mathrm{MinRTT_{curr\_rtt}}\le\gamma\times\mathrm{MinRTT_{before\_probe}}$, BBRv2+ re-enters ProbeUp. The rationale behind this is that there is possible more free bandwidth capacity as no significant increment of queuing delay arises in the current ProbeUp 
sub-state. 

In the case of bandwidth decrements, BBRv2+ needs to update its \textit{BtlBW} estimation to new bandwidth measurements as soon as possible. When bandwidth decreases, BBRv2+ sends data faster than the bottleneck bandwidth and packets accumulate in the buffer of the bottleneck link. We thus also leverage the \signame{} to detect bandwidth decrement. If BBRv2+ is in ProbeCruise or ProbeDown, on the receipt of a new ACK in an RTT round, Algorithm~\ref{alg:BBRv2p_advance_filter} is called\footnote{It is called once at maximum for every RTT round to avoid expiring the \textit{BtlBW} estimation too frequently.}. The reason that the algorithm is only applicable in ProbeCruise or ProbeDown is to eliminate the impact on delay variations caused by ProbeTry and ProbeUp sub-states. In Algorithm~\ref{alg:BBRv2p_advance_filter}, BBRv2+ expires its current \textit{BtlBW} estimation if $\mathrm{MinRTT_{curr\_rtt}}$ is larger than the recently measured minimum RTT by $\theta$ times. $\theta$ is a parameter to balance the speed to converge to new bandwidth and the resistance to noises in bandwidth measurements. A small $\theta$ may lead to throughput oscillation, while a large $\theta$ may reduce BBRv2+'s responsiveness to bandwidth dynamics. We recommend $\theta\in[1.05,1.15]$ according to our experiences.

\begin{algorithm}[htb]
    \caption{Advance \textit{BtlBW} max\_filter}
    \label{alg:BBRv2p_advance_filter}
    \LinesNumbered
    \SetKwProg{Fn}{Function}{}{end}
    \SetKwFunction{expire}{expire\_the\_oldest\_value}
    \SetKwInOut{Input}{Input}
    \SetKwInOut{Output}{Output}
    \SetKw{return}{return}
    \SetKw{true}{true}
    \SetKw{false}{false}
    \SetKw{not}{not}
    \SetKw{and}{and}
    \SetKw{or}{or}
    \Input{\texttt{conn}: BBRv2+ TCP connection}
    
    
    
    
    target\_rtt $\xleftarrow{}$ $\theta$ $*$ \texttt{conn}.\textit{RTprop}
    
    should\_advance $\xleftarrow{}$ (\texttt{conn}.$\mathrm{MinRTT_{curr\_rtt}}$ $>$ target\_rtt)


    \uIf{should\_advance}{\expire{{\normalfont\ttfamily conn}.\textit{BtlBW}}}
\end{algorithm}

\subsection{Dual-mode mechanism}
\label{sec:design:coexist}

Due to the use of the \signame{} to guide the aggressiveness in bandwidth probing, BBRv2+ will be starved by loss-based CCAs under deep buffers, suffering from the similar problem existing in most delay-based CCAs~\cite{al-saadi_survey_2019}. The root cause is that loss-based CCAs constantly fill the bottleneck buffer, where BBRv2+ falsely treats the increments of RTT as the signal of bandwidth decrements.

As BBRv2+ periodically drains the bottleneck buffer, during which the measured minimum RTT ($\mathrm{MinRTT_{curr\_cruise}}$ listed in Table.~\ref{tab:bbrv2plus_vars}) is close to \textit{RTprop} if no loss-based competitor exists. By comparing the $\mathrm{MinRTT_{curr\_cruise}}$ with the recorded \textit{RTprop} value, BBRv2+ estimates the existence of loss-based competitors. If loss-based competitors co-exist, BBRv2+ switches to use BBRv2's ProbeBW state, which enables BBRv2+ to co-exist with loss-based CCAs in the same way as BBRv2 that does not yield up obtained bandwidth due to RTT increments. Further, if the loss-based competitors no longer exist, BBRv2+ returns back to use the redesigned ProbeBW state. We note that the dual-mode mechanism does not switch BBRv2+ to use BBRv2's ProbeBW state if the bottleneck buffer is very shallow, because loss-based CCAs can not bloat the bottleneck buffer and BBRv2+ will not be starved.



\begin{algorithm}[htb]
    \caption{The dual-mode mechanism}
    \label{alg:BBRv2p_switch}
    \LinesNumbered
    \SetKwProg{Fn}{Function}{}{end}
    \SetKwFunction{maxfilter}{max\_filter}
    \SetKwInOut{Input}{Input}
    \SetKwInOut{Output}{Output}
    \SetKw{return}{return}
    \SetKw{true}{true}
    \SetKw{false}{false}
    \SetKw{not}{not}
    \SetKw{and}{and}
    \SetKw{or}{or}
    \SetKw{return}{return}
    \SetKwFunction{restart}{restart\_from\_startup}
    \Input{\texttt{conn}: BBRv2+ TCP connection}
    
    
    
    
    \uIf{{\normalfont\ttfamily conn}.probe\_bw\_mode $=$ BBRv2{\normalfont+}}{ \label{alg:mode:bbrv2p_start}
        switch\_thld $\leftarrow$ $\lambda_{1}$ $*$ \texttt{conn}.$\mathrm{\textit{RTprop}}$
        
        \uIf{{\normalfont\ttfamily conn}.$MinRTT_{curr\_cruise}$ $>$ switch\_thld}{
            \texttt{conn}.buffer\_filling$++$
        }
        \uElse{
            \texttt{conn}.buffer\_filling $\leftarrow$ 0
        }
        \uIf{{\normalfont\ttfamily conn}.buffer\_filling $\geq$ $\eta_{1}$}{
            \texttt{conn}.probe\_bw\_mode $\leftarrow$ BBRv2
            
            \restart{{\normalfont\ttfamily conn}} \label{alg:mode:bbrv2p_end}
        }
    }
    \uElse{
        switch\_thld $\leftarrow$ $\lambda_{2}$ $*$ \texttt{conn}.$\mathrm{\textit{RTprop}}$
        
        \uIf{{\normalfont\ttfamily conn}.$MinRTT_{curr\_cruise}$ $\leq$ switch\_thld}{
            \texttt{conn}.buffer\_empty$++$
        }
        \uElse{
            \texttt{conn}.buffer\_empty $\leftarrow$ 0
        }
        \uIf{{\normalfont\ttfamily conn}.buffer\_empty $\geq$ $\eta_{2}$}{
            \texttt{conn}.probe\_bw\_mode $\leftarrow$ BBRv2+
        }
    }
\end{algorithm}


The dual-mode mechanism is detailed in Algorithm~\ref{alg:BBRv2p_switch}, which runs at the end of ProbeCruise. If the sender is running in BBRv2+'s ProbeBW state and has not seen RTT samples close to the \textit{RTprop} for a number of ($\eta_{1}$) successive ProbeCruise sub-states, it switches to use BBRv2's ProbeBW state and restarts itself from Startup (line~\ref{alg:mode:bbrv2p_start}--\ref{alg:mode:bbrv2p_end}). We note that \texttt{restart\_from\_startup(conn)} in line~\ref{alg:mode:bbrv2p_end} is a heuristic to quickly regain the bandwidth that has been potentially yielded up to loss-based competitors by BBRv2+ recently. If the sender's ProbeBW state is BBRv2 and it has seen low RTTs for $\eta_{2}$ successive ProbeCruise sub-states, it returns back to use BBRv2+'s ProbeBW state because the competitors are most likely gone. We note that the four parameters in Algorithm~\ref{alg:BBRv2p_switch}, $\lambda_{1}$, $\lambda_{2}$, $\eta_{1}$, and $\eta_{2}$, are to control the sensitivity of BBRv2+ to the co-existence of loss-based CCAs. In practice, we used 1.1, 1.05, 2, and 4 for $\lambda_{1}$, $\lambda_{2}$, $\eta_{1}$, and $\eta_{2}$ respectively. Nevertheless, these parameters can be tuned to fit specific networks in user space in our current implementation. 

\subsection{Compensation for BDP estimation}
\label{sec:design:comp_cwnd}

We have seen in \S\ref{sec:measurement:jitter}, when network jitters are high, BBRv2 (also BBR) underestimates \textit{RTprop}, thus the BDP of the network path, leading to \textit{cwnd} exhaustion and thus performance degradation. To boost BBRv2+'s performance under high network jitters, BBRv2+ takes network jitters into account when estimating the BDP of the network path.

BBRv2+ compensates the BDP estimation with a component proportional to RTT variations when network jitters are high, which is detailed in Algorithm.~\ref{alg:BBRv2p_comp_cwnd}. As instantaneous RTT variations could be very dynamic, to ensure that BBRv2+ can tolerate jitters up to the maximum extent, we use the recently measured maximum RTT variation, $\mathrm{Max_{4RTT}(jitter)}$ in Table.~\ref{tab:bbrv2plus_vars}, as the indicator of recent jitters. When $\mathrm{Max_{4RTT}(jitter)}$ exceeds $\mu\times\mathrm{\textit{RTprop}}$, the estimated \textit{RTprop} is increased to the sum of the original \textit{RTprop} and the delay variation ($\mathrm{Max_{4RTT}(jitter)}$) to mitigate the underestimation of \textit{RTprop}. We recommend setting $\mu$ around 0.5 because the performance of BBRv2 starts to degrade when jitters approach half of \textit{RTprop} as observed in \S\ref{sec:measurement:jitter}. 

\begin{algorithm}[htb]
    \caption{Compensating BDP estimation}
    \label{alg:BBRv2p_comp_cwnd}
    \LinesNumbered
    \SetKwProg{Fn}{Function}{}{end}
    \SetKwFunction{maxfilter}{max\_filter}
    \SetKwInOut{Input}{Input}
    \SetKwInOut{Output}{Output}
    \SetKw{return}{return}
    \Input{\texttt{conn}: BBRv2+ TCP connection}
    \Output{the BDP estimation of BBRv2+}
    
    
    jitter $\xleftarrow{}$ \texttt{conn}.$\mathrm{Max_{4RTT}(jitter)}$
    
    threshold $\xleftarrow{}$ $\mu$ $*$ \texttt{conn}.\textit{RTprop}
    
    fixed\_\textit{RTprop} $\leftarrow$ \texttt{conn}.\textit{RTprop}

    \uIf{jitter $>$ threshold}{fixed\_\textit{RTprop} $\leftarrow$ fixed\_\textit{RTprop} $+$ jitter}
    \return \texttt{conn}.\textit{BtlBW} $*$ fixed\_\textit{RTprop}
    
\end{algorithm}




\subsection{Implementation}
\label{sec:design:impl}

BBRv2+ is implemented as a Linux kernel module ($\sim$2100 LoCs), based on Google's BBRv2 alpha kernel module~\cite{bbr2_kernel}. Therefore, it is easy to deploy BBRv2+ on the hosts where BBRv2 is already in use. The parameters of BBRv2+ are exposed to user-space through the \texttt{/sys/module} interfaces, which allows users to change the parameters according to their need without recompiling the kernel module. The code of BBRv2+ is open-sourced on Github~\cite{furongYangfurongBBRv2plus2021} to the research community for further test and improvement. 


%


%% file: sections/BBRv2p-evaluation.tex
\label{sec:eva}

In this section, we evaluate BBRv2+ based on both Mininet-based emulation and real-world trace driven emulation. First, we describe our experiment setup in \S\ref{sec:eva:method}. We then evaluate the benefits of BBRv2+ from the perspectives of the responsiveness to bandwidth changes (\S\ref{sec:eva:bw_change}) and the resilience to network jitters (\S\ref{sec:eva:jitter}). Next, we demonstrate that BBRv2+ is able to keep the advantages of BBRv2 in inter-protocol fairness (\S\ref{sec:eva:inter_fair}),  RTT fairness (\S\ref{sec:eva:rtt_fair}), and low retransmissions (\S\ref{sec:eva:retrans}). Finally, we evaluate the performance of BBRv2+ through real-world trace driven emulation in \S\ref{sec:eva:mahimahi}.


\subsection{Evaluation setup}
\label{sec:eva:method}

\begin{figure}[ht]
    \centering
    \includegraphics[width=0.8\linewidth]{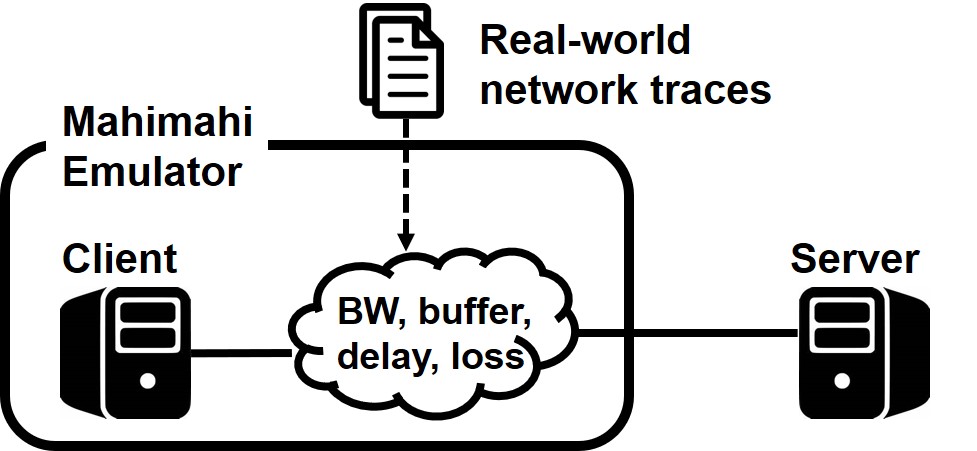}
    \caption{Mahimahi testbed}
    \label{fig:mahimahi_topo}
\end{figure}

Two testbeds are used for the evaluation of BBRv2+. One is the Mininet-based testbed used in \S\ref{sec:measurement}, as a controlled environment to evaluate BBRv2+ from various perspectives. The other is based on Mahimahi~\cite{netravali_mahimahi_2015}, a trace-driven emulator that can accurately replay real-world packet-level traces, as illustrated in Fig.~\ref{fig:mahimahi_topo}. The physical server running the two testbeds is the same one as that used to evaluate BBRv2 in \S\ref{sec:measurement}. The toolset for data analysis and traffic generation is also the same as that in \S\ref{sec:measurement}.


In all experiments in this section, the $\alpha$ (the loss threshold) of BBRv2+ is set to 20\% as the value is suitable for most of the buffer sizes according to the results in \S\ref{sec:measurement:loss_resilience}.


\subsection{Responsiveness to bandwidth dynamics}
\label{sec:eva:bw_change}

\begin{figure*}[ht]
    \begin{minipage}{0.45\linewidth}
        \centering
        \includegraphics[width=0.95\linewidth]{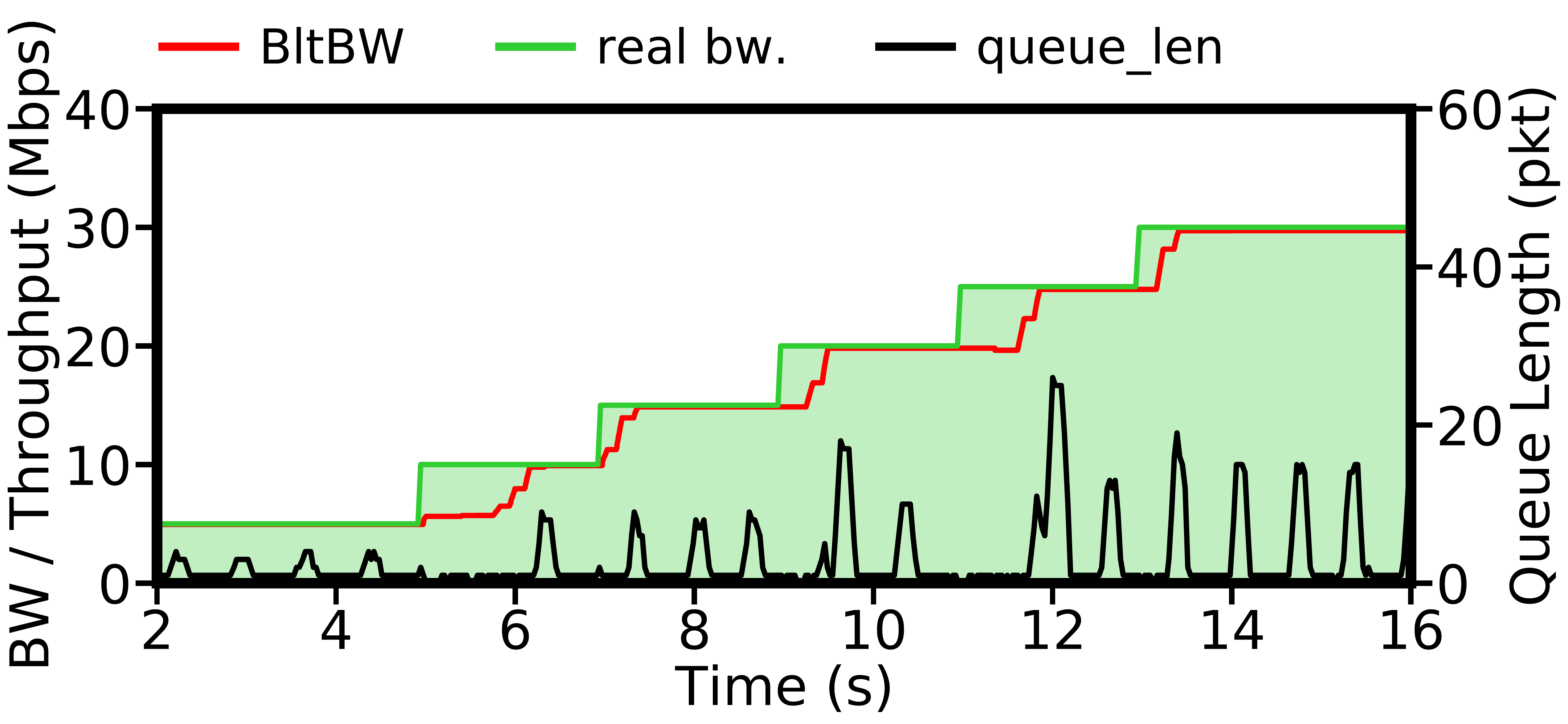}
        \subcaption{bw. increasing}
        \label{fig:bw_inc_bbr2p}
    \end{minipage}
    \hfill
    \begin{minipage}{0.45\linewidth}
        \centering
        \includegraphics[width=0.95\linewidth]{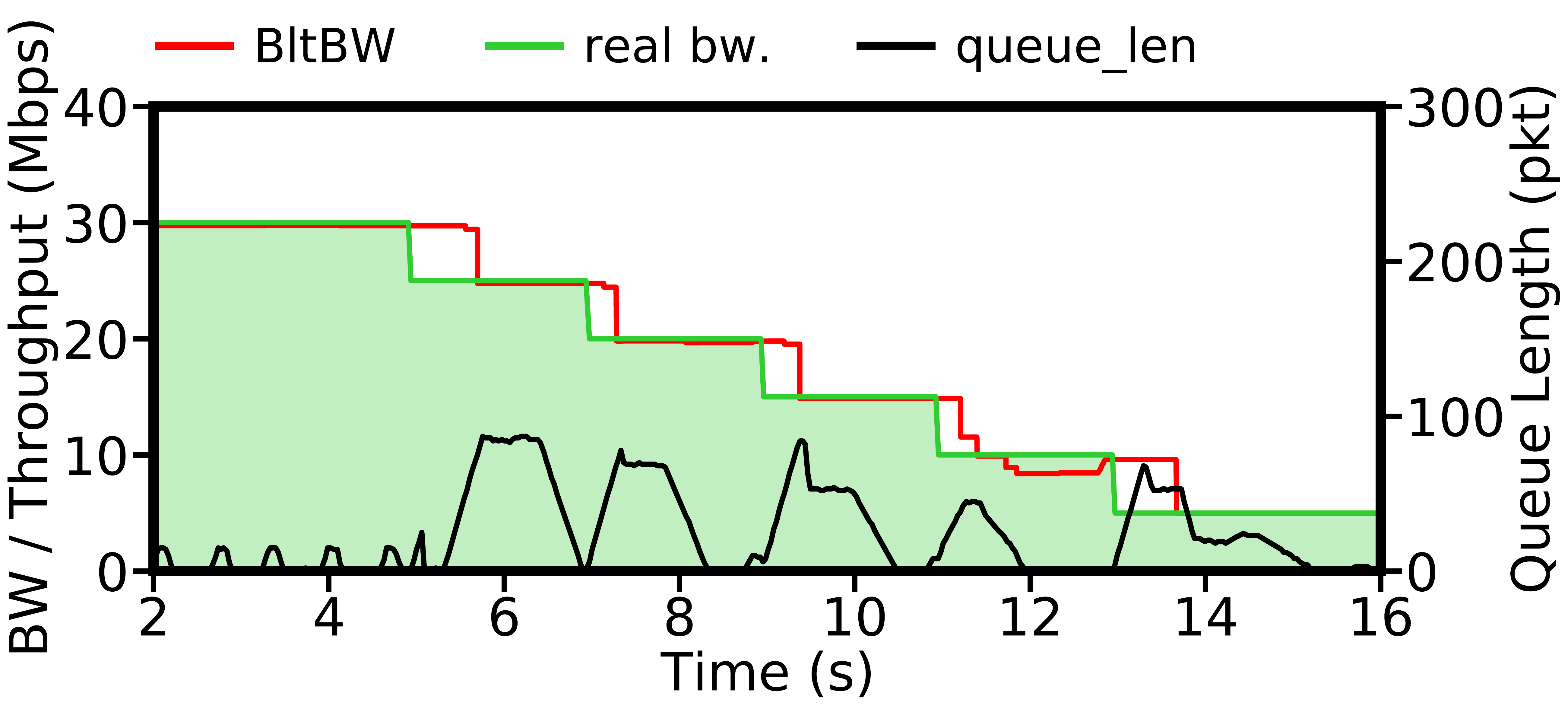}
        \subcaption{bw. decreasing}
        \label{fig:bw_dec_bbr2p}
    \end{minipage}
    \caption{BBRv2+'s responsiveness to bandwidth increases (a) and decreases (b). The red line represents the \textit{BtlBW} estimation of BBRv2+, and the green line indicates the real bandwidth of the bottleneck. The dark line shows the dynamics of the bottleneck link's queue length.}
    \label{fig:bw_change_bbr2p}
\end{figure*}

To evaluate BBRv2+'s responsiveness to bandwidth dynamics, we use the same settings as that in \S\ref{sec:measurement:bw_dynamics}, in order to run BBRv2+ to have a microscopic view on how it reacts to bandwidth dynamics. Fig.~\ref{fig:bw_change_bbr2p} shows the results.

In Fig.~\ref{fig:bw_inc_bbr2p}, when there is no bandwidth increment, BBRv2+ only enters ProbeTry for a very short duration and finishes bandwidth probing very soon, which leads to an instantaneous short standing queue. However, when the bandwidth is increased, BBRv2+ can timely adapt its' \textit{BtlBW} estimation to the real bandwidth. Compared with the results of BBR in Fig.~\ref{fig:bw_inc_bbr} and BBRv2 in Fig.~\ref{fig:bw_inc_bbr2}, BBRv2+ is capable to utilize newly available bandwidth as quick as BBR, while its guided probing strategy (by the \signame{}) incurs lower queuing delay than BBR.

In Fig.~\ref{fig:bw_dec_bbr2p}, when bandwidth decreases, BBRv2+ notices that the queuing delay is obviously rising up via increased RTT (see \S\ref{sec:design:probebw}). It expires the old  \textit{BtlBW} estimation and adapts its  \textit{BtlBW} estimation to the available bandwidth. Compared with the results of BBR and BBRv2 in Fig.~\ref{fig:resp_to_bw_dynamics}, BBRv2+ adapts its sending rate to the decreased bandwidth much faster, which leads to lower queuing delay.

\begin{figure}[htb]
    \centering
    \includegraphics[width=0.7\linewidth]{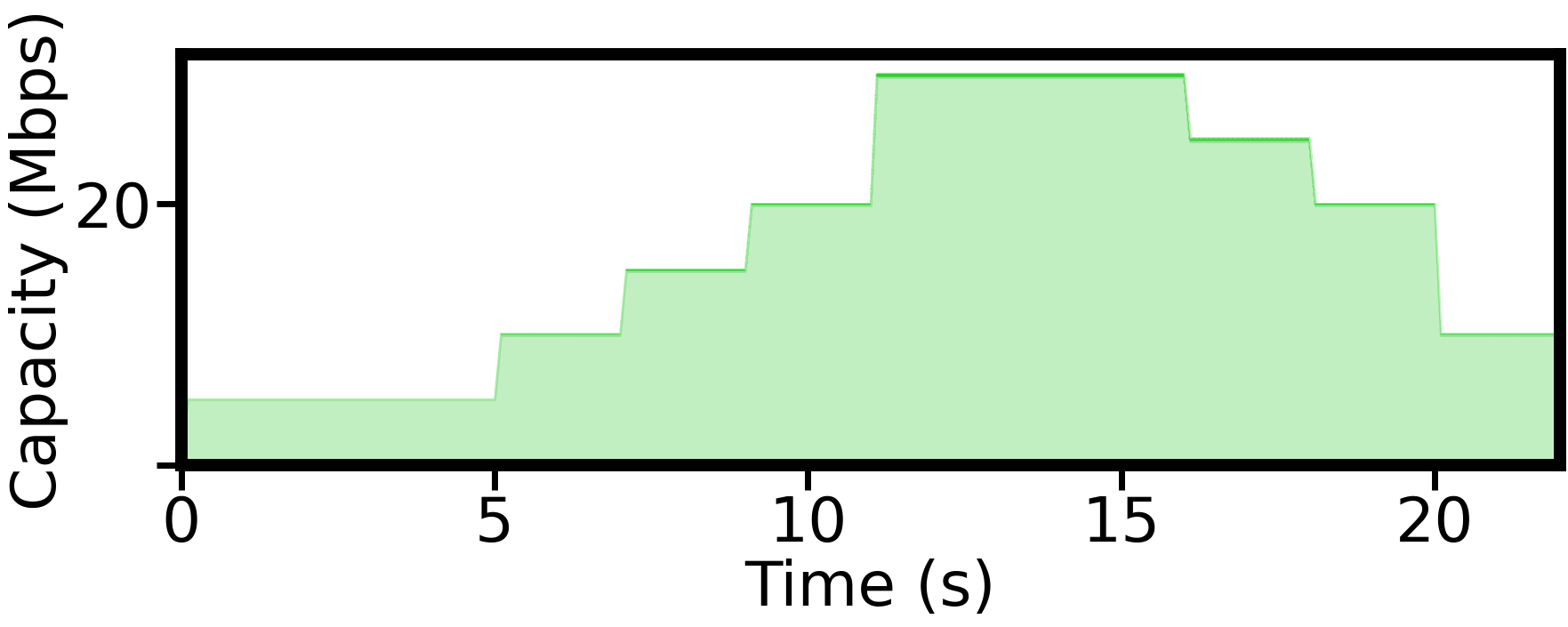}
    \caption{An example of traces with bandwidth changing as a step function.}
    \label{fig:step_bw_trace}
\end{figure}

Next, we compare the responsiveness of Cubic, BBR, BBRv2, and BBRv2+ to bandwidth dynamics in our trace-driven emulation testbed Mahimahi, using five synthesized network traces where the bandwidth changes as step functions, as illustrated in Fig.~\ref{fig:step_bw_trace}. Following the settings in~\cite{abbasloo_classic_2020}, we set the buffer size to 1.5MB, the delay to 20ms, and the loss rate to zero. In each experiment, the flow throughput, as well as the sojourn time of each packet in the buffer of the bottleneck link (denoted as queuing delay), are recorded. 


\begin{figure}[htb]
    \centering
    \includegraphics[width=0.7\linewidth]{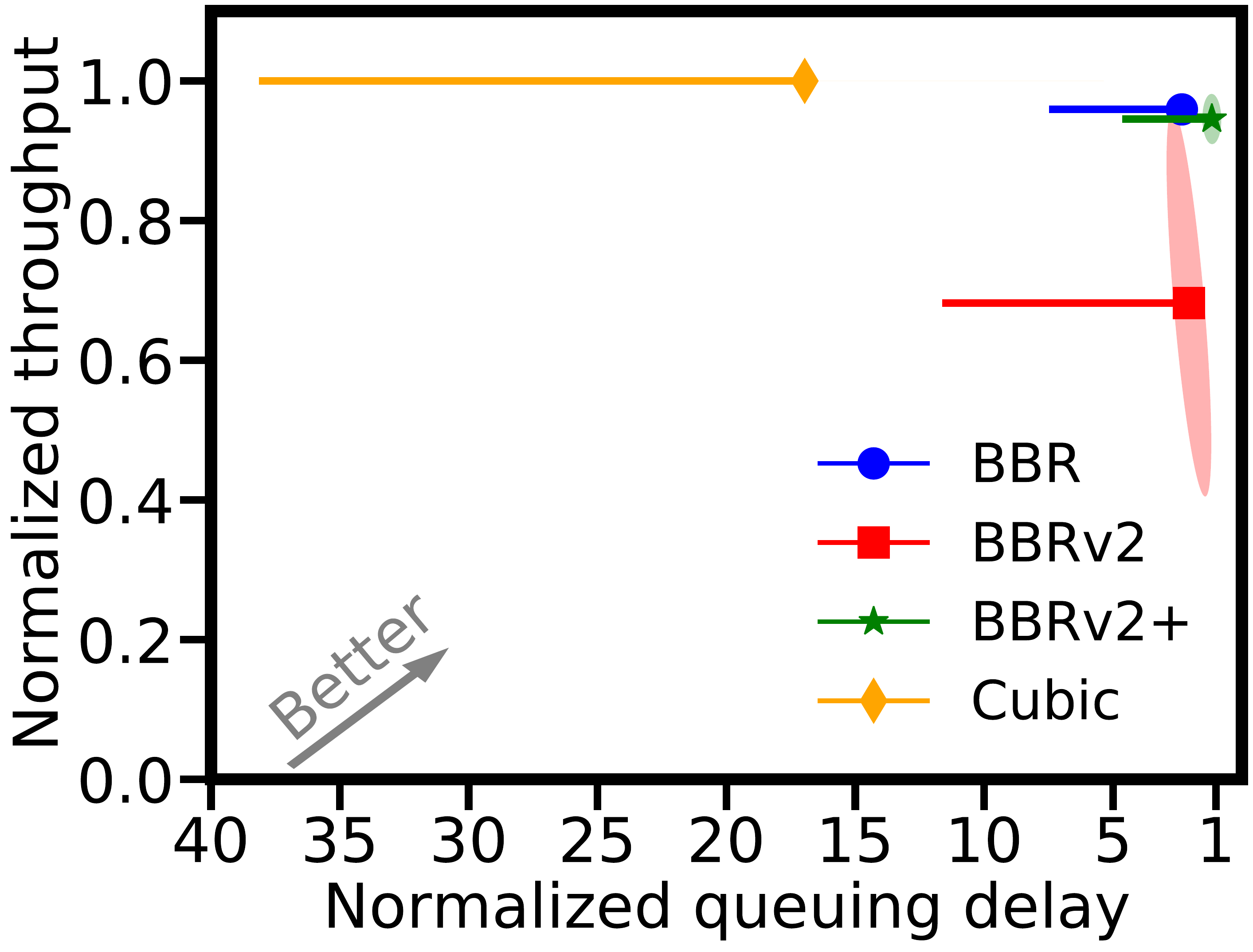}
    \caption{Normalized throughput and queuing delay of different CCAs. (markers: average throughput and queuing delay; left end of the lines: 95\%-tile of queuing delay; ellipses: the standard deviations)}
    \label{fig:step_bw_trace_res}
\end{figure}

To compare the overall performance of all CCAs on a network trace, we normalized the average queuing delay and the average throughput of all CCAs to the minimum average queuing delay and the maximum average throughput achieved on that trace, respectively. In addition, we also normalized the 95\%tile queuing delay of all CCA on a network trace to the minimum average queuing delay achieved on that trace. Then, we averaged all normalized values over all traces.  The results are shown in Fig.~\ref{fig:step_bw_trace_res}. 
We observe that BBRv2+ achieved significantly higher throughput and lower queuing delay than BBRv2, and lower queuing delay at the cost of slightly lower throughput than BBR. These observations stem from the facts that: \textbf{(1)} BBRv2+ probes for bandwidth at a frequency similar to BBR's one, thus, achieving high bandwidth utilization as BBR does; \textbf{(2)} BBRv2+ adapts its sending rate to decreased bandwidth faster than BBR as it quickly updates its \textit{BtlBW} estimation upon increased queuing delay.


\subsection{Resilience to network jitters}
\label{sec:eva:jitter}

\begin{figure}[ht]
    \begin{minipage}{0.48\linewidth}
        \centering
        \includegraphics[width=1\linewidth]{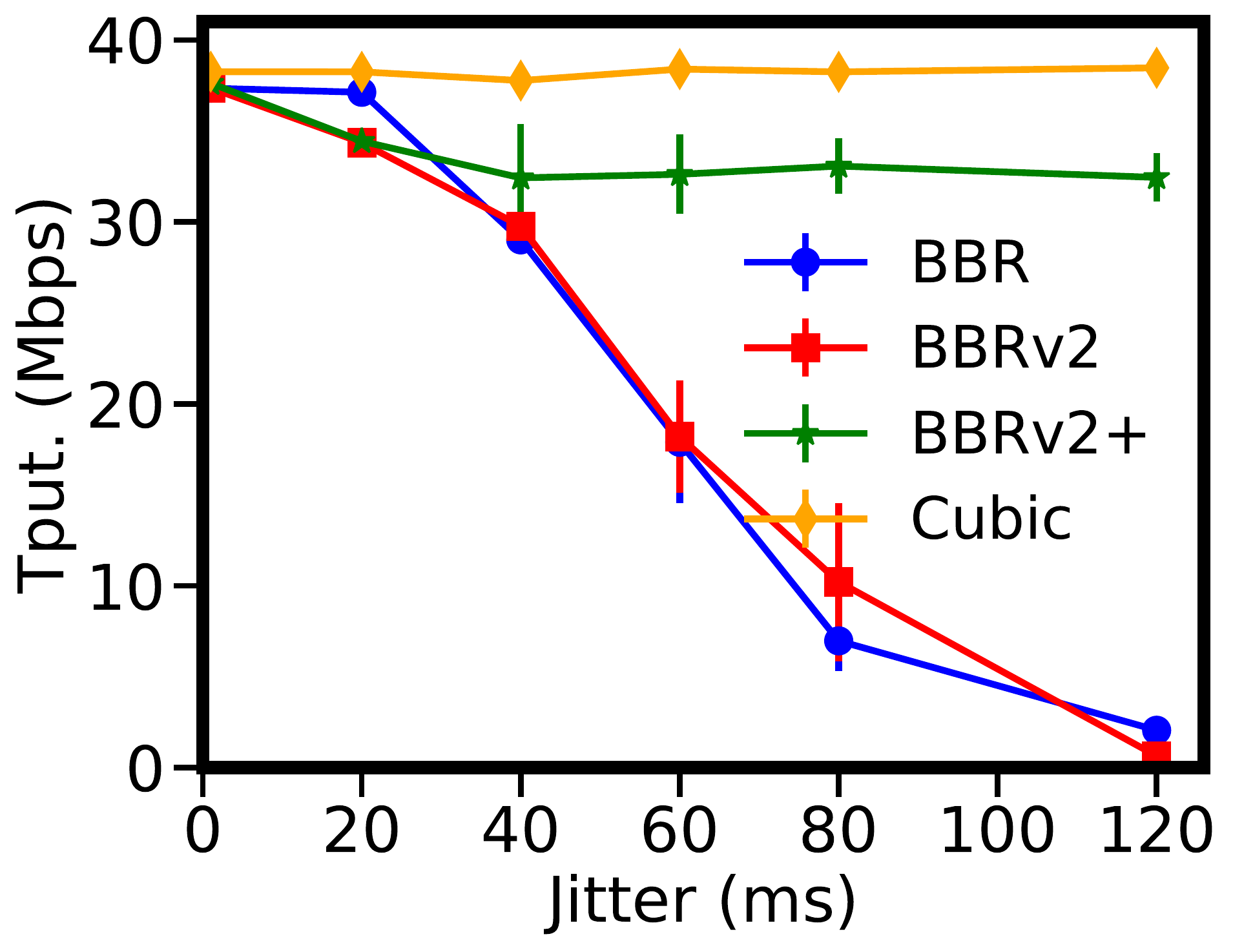}
        \subcaption{Throughput}
        \label{fig:bbr2p_tput_vs_jitters}
    \end{minipage}
    \hfill
    \begin{minipage}{0.48\linewidth}
        \centering
        \includegraphics[width=1\linewidth]{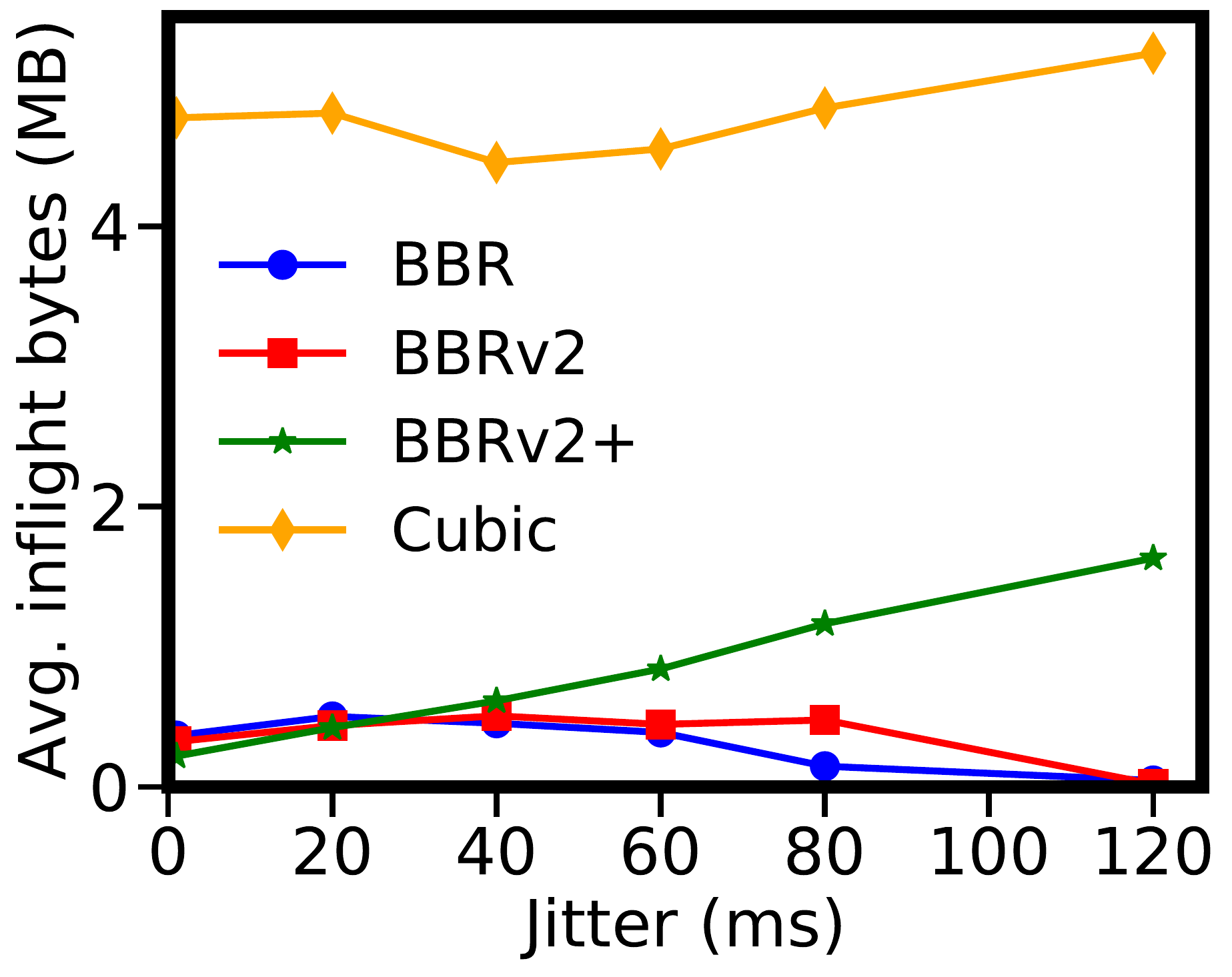}
        \subcaption{Avg. inflight bytes}
        \label{fig:bbr2p_owin}
    \end{minipage}
    \caption{Resilience to network jitters}
    \label{fig:bbr2p_resilience_to_jitters}
\end{figure}

Next, we evaluate the performance of BBRv2+ under network jitters, using the same settings as in \S\ref{sec:measurement:jitter}. The throughput of BBRv2+, Cubic, BBR and BBRv2 under various levels of jitters are shown in Fig.~\ref{fig:bbr2p_tput_vs_jitters}. Different from BBR and BBRv2, the throughput of BBRv2+ does not degrade when the network jitters become larger. Fig.~\ref{fig:bbr2p_owin} further plots the average inflight bytes of four CCAs; the results confirm that the BDP compensation mechanism of BBRv2+ succeeds to increase the inflight size for higher throughput when the network jitter becomes larger. Nevertheless, the throughput of BBRv2+ is slightly lower than Cubic. The reason is that our compensation to BDP is a bit conservative; in contrast, Cubic's \textit{cwnd} can grow far beyond the real BDP as it is not affected by network jitters. 

\subsection{Inter-protocol fairness}
\label{sec:eva:inter_fair}

\begin{figure}[ht]
    \centering
    \begin{minipage}{0.48\linewidth}
        \centering
        \includegraphics[width=1\linewidth]{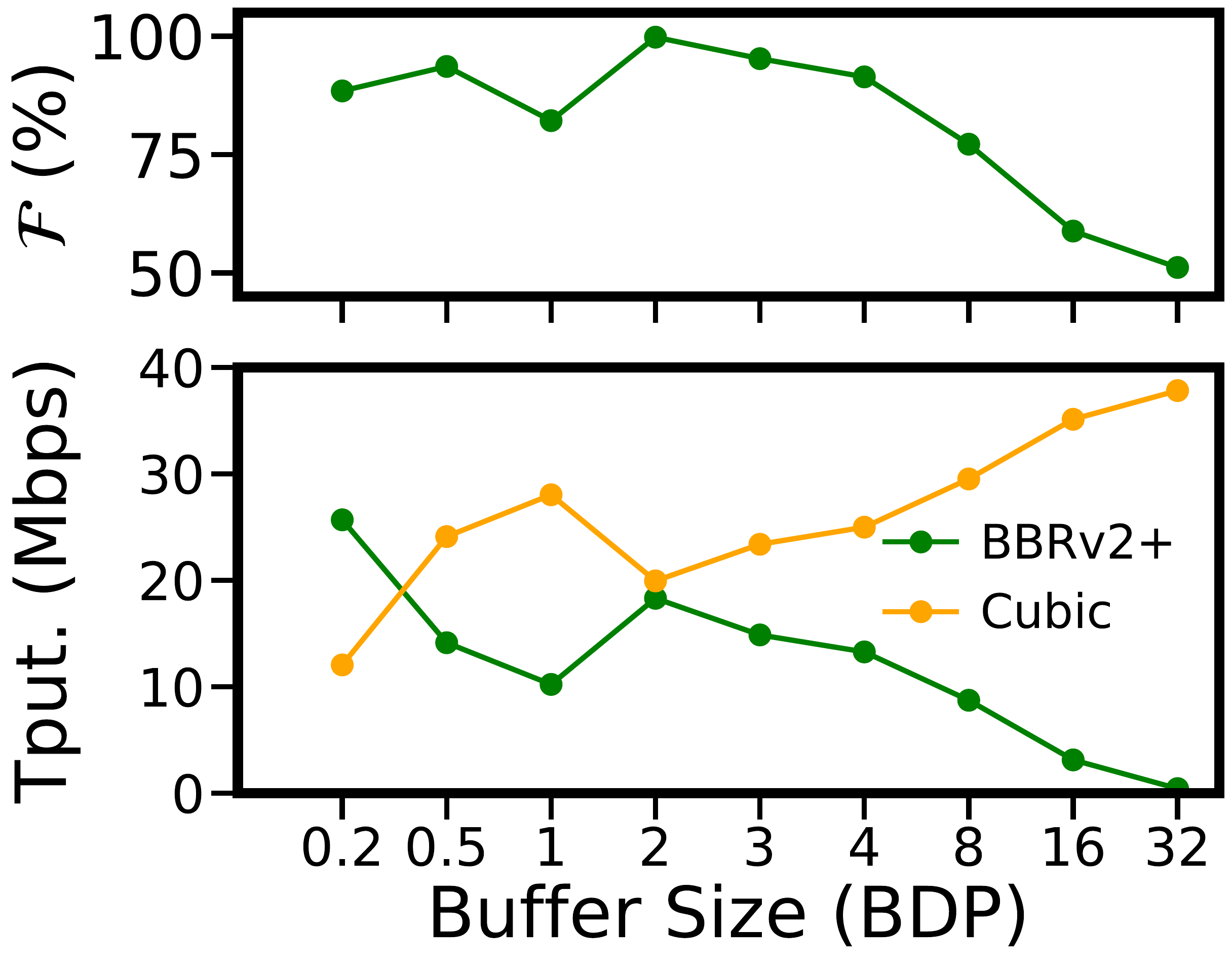}
        \subcaption{Without the dual-mode mechanism}
        \label{fig:cc_fair_bbr2p_wo_fallback}
    \end{minipage}
    \hfill
    \begin{minipage}{0.48\linewidth}
        \centering
        \includegraphics[width=1\linewidth]{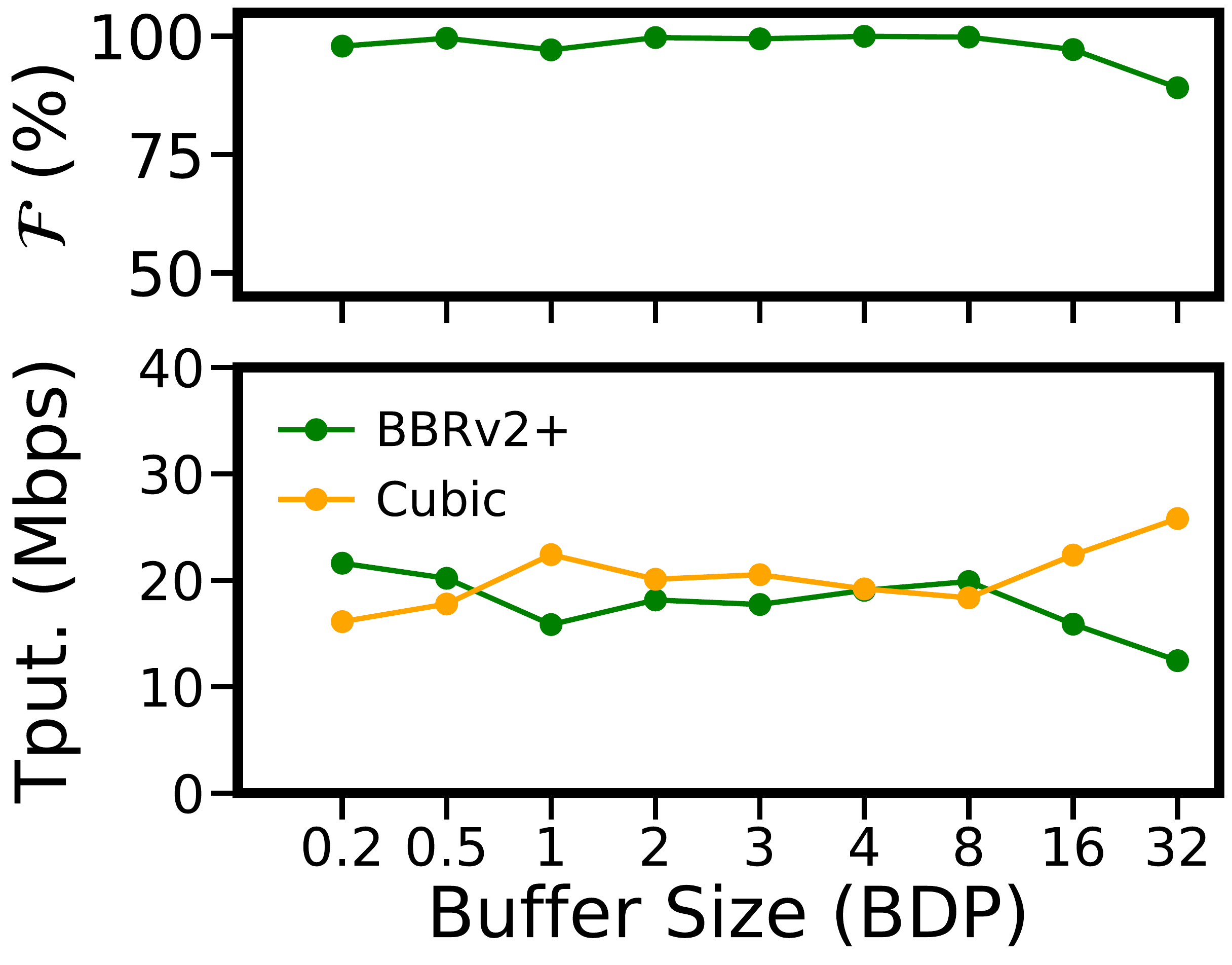}
        \subcaption{With the dual-mode mechanism}
        \label{fig:cc_fair_bbr2p_w_fallback}
    \end{minipage}
    \caption{BBRv2+: Inter-protocol fairness}
    \label{fig:cc_fair_bbr2p}
\end{figure}

The inter-protocol fairness of BBRv2+ is evaluated using the same settings as that in \S\ref{sec:measurement:inter_fair}. We considered BBRv2+ with/without the dual-mode mechanism (see \S\ref{sec:design:coexist}) to study the impact of this mechanism on inter-protocol fairness. The results are shown in Fig.~\ref{fig:cc_fair_bbr2p}. Several observations are notable.

First, the results demonstrate the efficacy of the dual-mode mechanism. In Fig.~\ref{fig:cc_fair_bbr2p_wo_fallback}, we can observe that BBRv2+ without the dual-mode mechanism is starved by Cubic in deep-buffered cases. This is because BBRv2+ falsely treats the RTT increments caused by Cubic as a signal of bandwidth shrinking, thus, constantly yielding up bandwidth to Cubic. The problem is eliminated by the dual-mode mechanism as shown in Fig.~\ref{fig:cc_fair_bbr2p_w_fallback}.

Second, compared with the results of BBRv2(20\%, 0.3) in Fig.~\ref{fig:cc_fair_bbr2_20alpha}, we can see that: \textbf{(1)} BBRv2+ provides better inter-protocol fairness than BBRv2(20\%, 0.3) under an extremely shallow buffer (i.e. 0.2$\times$ BDP); \textbf{(2)} BBRv2+'s inter-protocol fairness is similar to that of BBRv2(20\%, 0.3) under other buffer sizes. The reason for the better inter-protocol fairness of BBRv2+ under an extremely shallow buffer is that BBRv2+ does not enter ProbeUp, which is more aggressive than ProbeTry, thanks to the two-step probing mechanism (see \S\ref{sec:design:probebw}) while BBRv2(20\%, 0.3) periodically enters ProbeUp. 

Third, compared with the results of BBR in Fig.~\ref{fig:cc_fairness_bbr} and BBRv2 in Fig.~\ref{fig:cc_fairness_bbr2}, BBRv2+ performs no worse than the better one among BBR and BBRv2 under different buffer sizes. The reasons are three-fold: \textbf{(1)} under shallow buffers, BBRv2+ achieves similar inter-protocol fairness to that of BBRv2 thanks to its cautiously aggressive bandwidth probing strategy (see \S\ref{sec:design:probebw}); \textbf{(2)} under moderate buffers, BBRv2+ is close to BBRv2(20\%, 0.3) that has better inter-protocol fairness than BBRv2 in these cases as explained in \S\ref{sec:measurement:loss_resilience}; \textbf{(3)} under deep buffers, the three CCAs perform closely as they all have an inflight cap around 2$\times$ BDP, thus, unable to beat loss-based CCAs.


\begin{figure}[ht]
    \centering
    \begin{minipage}{0.5\linewidth}
        \centering
        \includegraphics[width=1\linewidth]{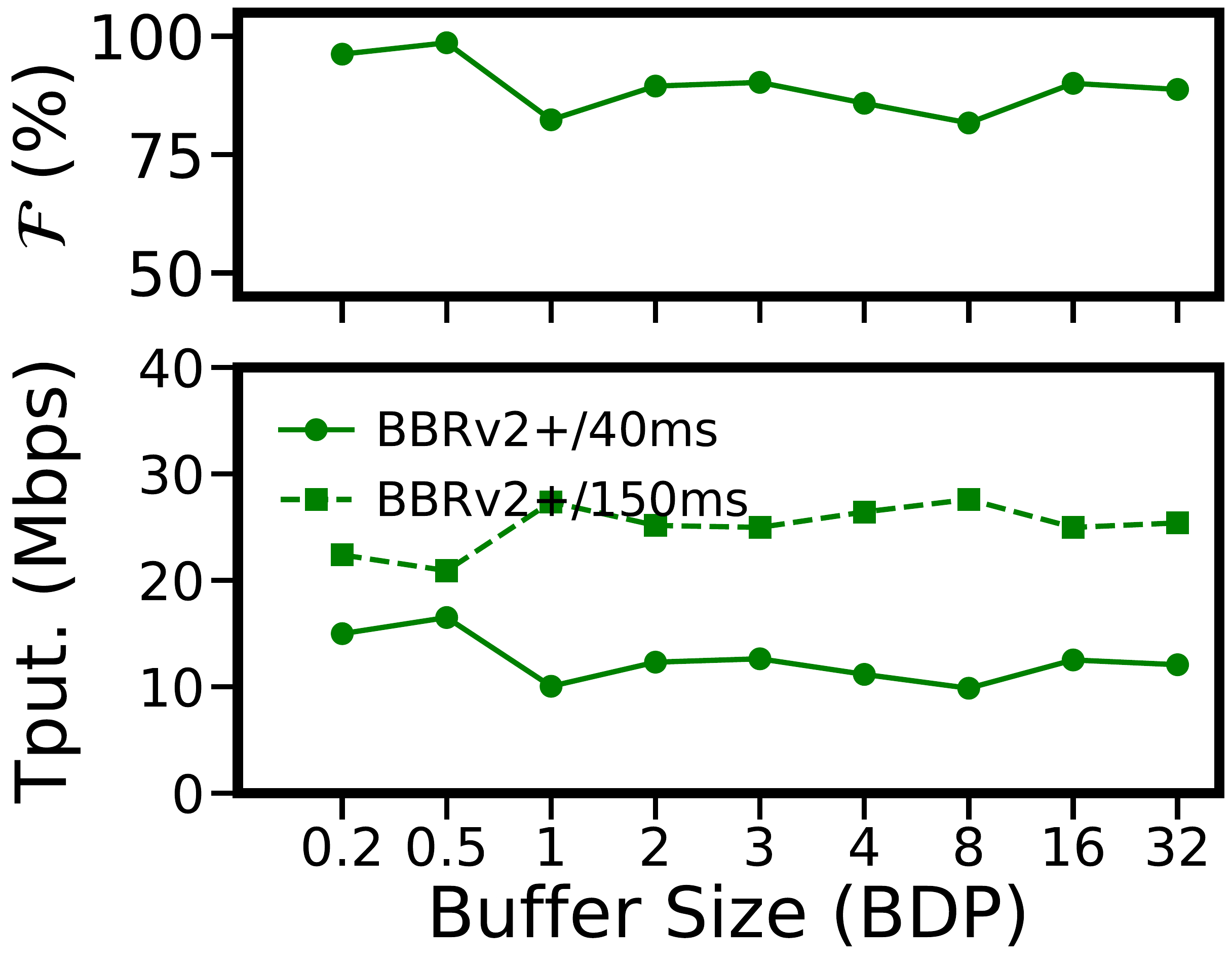}
    \end{minipage}
    \caption{BBRv2+: RTT fairness}
    \label{fig:rtt_fair_bbr2p}
\end{figure}



\subsection{RTT fairness}
\label{sec:eva:rtt_fair}

Next, we evaluate the RTT fairness of BBRv2+ using the same setting as that in \S\ref{sec:measurement:rtt_fair}. The results are presented in Fig.~\ref{fig:rtt_fair_bbr2p}. Compared with the results of BBR in Fig.~\ref{fig:rtt_fairness_bbr} and those of BBRv2 in Fig.~\ref{fig:rtt_fairness_bbr2}, BBRv2+ has better RTT fairness than BBR and behaves close to BBRv2.  The results are expected as the mechanisms in BBRv2 that improves the RTT fairness over BBR (see \S\ref{sec:measurement:rtt_fair}) remain unchanged in BBRv2+.


\subsection{Retransmissions in shallow-buffered networks}
\label{sec:eva:retrans}

\begin{figure}[htb]
    \centering
    \begin{minipage}{0.6\linewidth}
        \centering
        \includegraphics[width=1\linewidth]{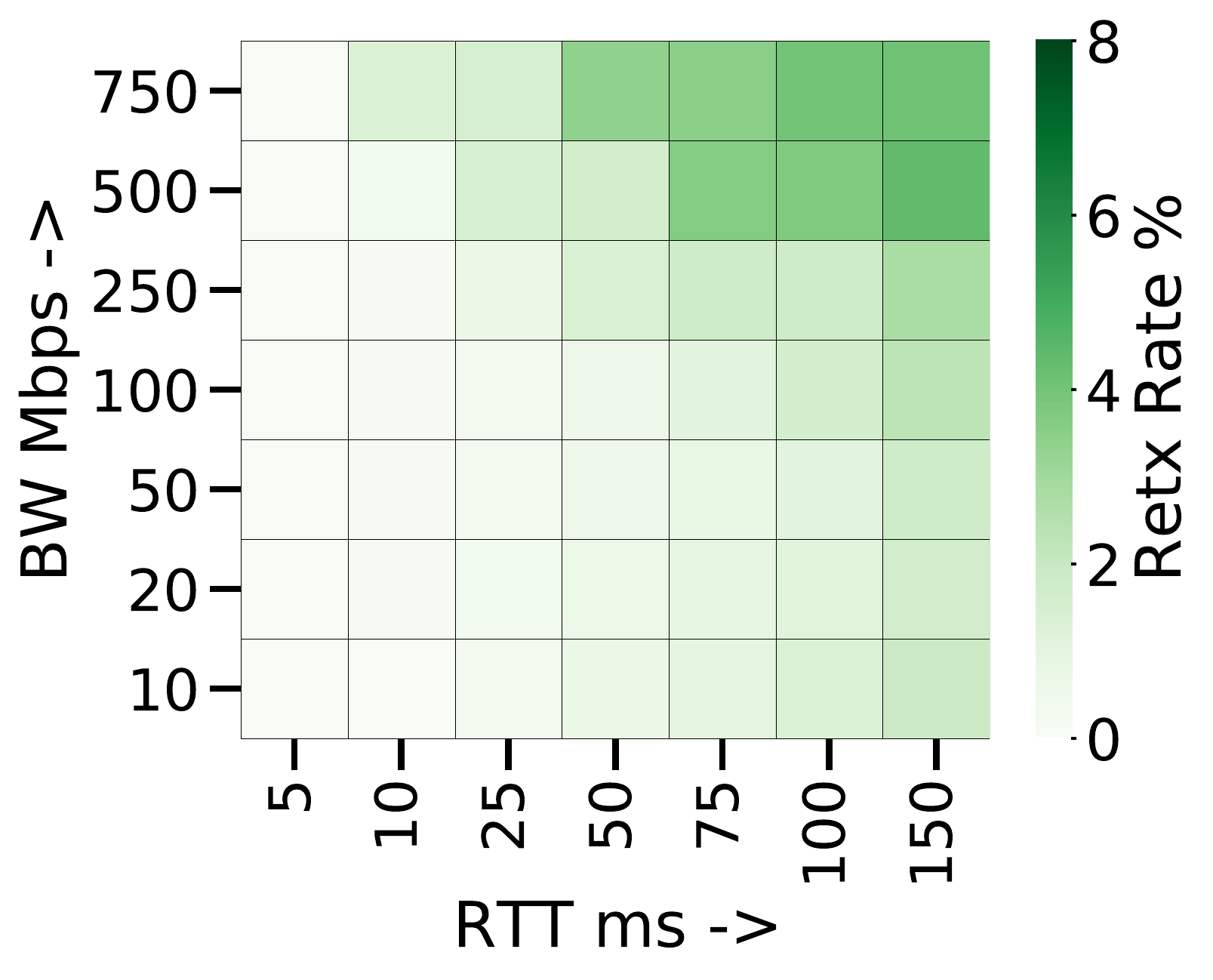}
    \end{minipage}
    \caption{BBRv2+: retransmission rate (100KB buffer)}
    \label{fig:bbr2p_vs_bbr2}
\end{figure}

In the following, we evaluate whether BBRv2+ is as aggressive as BBR to lead to excessive retransmissions in shallow-buffered networks, using the same setting as that in \S\ref{sec:retx_vs_tput}. The results are shown in Fig.~\ref{fig:bbr2p_vs_bbr2}. We observe that when the buffer is extremely shallow (e.g. 0.02$\times$ BDP when the bandwidth is 500Mbps and the RTT is 75ms), BBRv2+ incurs more retransmissions than BBRv2. This is because the bandwidth probing frequency of BBRv2+ is higher than that of BBRv2. Although BBRv2+ uses a relatively small \emph{pacing\_gain} (1.1) when it starts to probe for more bandwidth in ProbeTry, it still causes buffer overflow when the network buffer is extremely shallow. However, compared with the results of BBR in Fig.~\ref{fig:retx_bbr}, BBRv2+ reduces retransmissions significantly.


\subsection{Real-world trace driven emulation}
\label{sec:eva:mahimahi}

To evaluate how BBRv2+ performs in real network conditions, we compare BBRv2+ with Cubic, BBR, BBRv2, and Orca~\cite{abbasloo_classic_2020} in the emulation-based Mahimahi testbed, using traces collected in real-world networks. Orca\footnote{We directly used the model trained by the authors in our experiments.} is used for comparison as a representative of the state-of-the-art learning-based CCAs.

\begin{figure}[htb]
    \begin{minipage}{0.48\linewidth}
        \centering
        \includegraphics[width=1\linewidth]{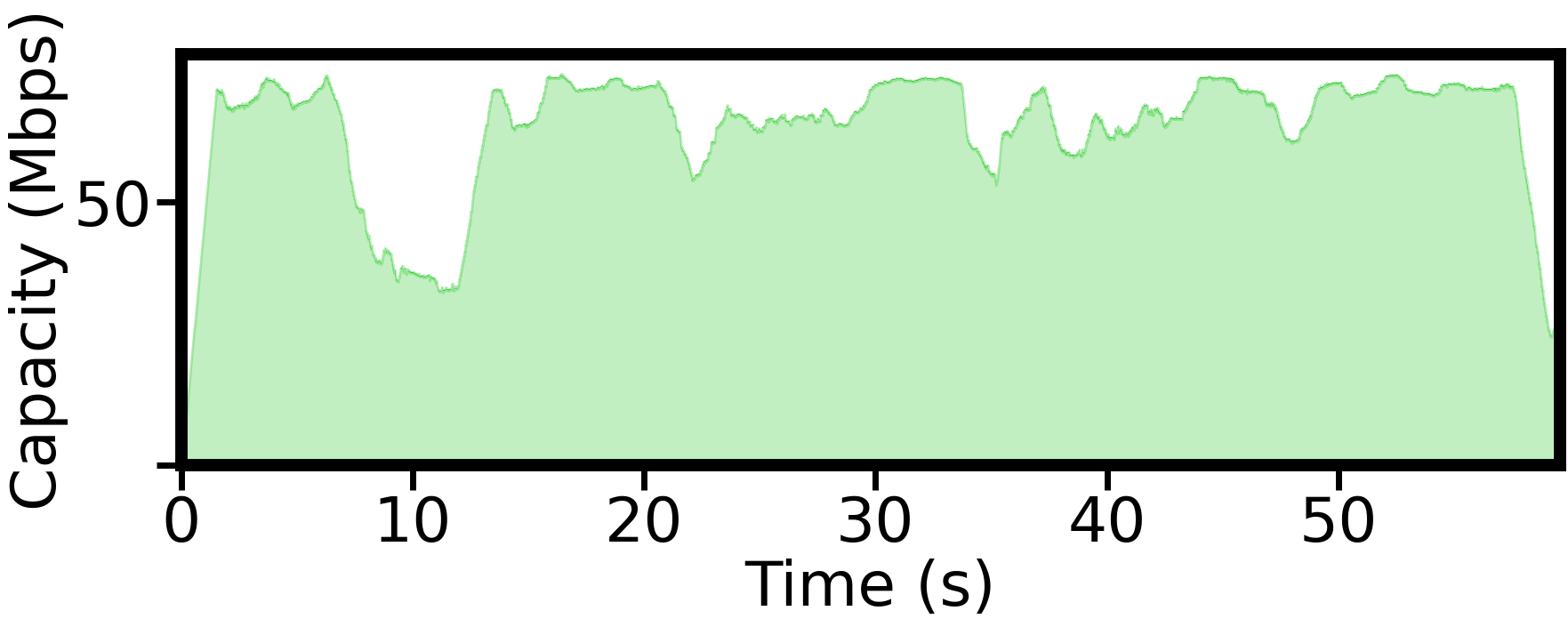}
        \subcaption{An example of stationary traces}
        \label{fig:mahimahi_stationary_trace}
    \end{minipage}
    \hfill
    \begin{minipage}{0.48\linewidth}
        \centering
        \includegraphics[width=1\linewidth]{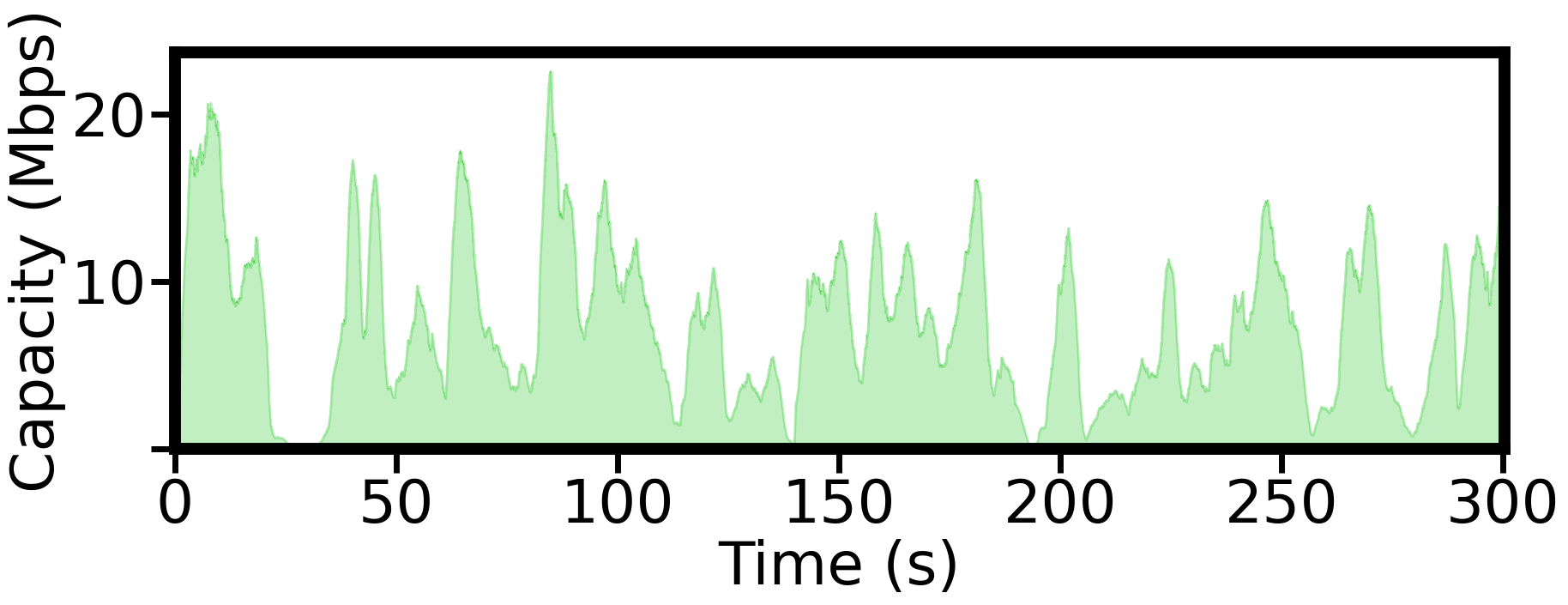}
        \subcaption{An example of high-mobility traces}
        \label{fig:mahimahi_mobility_trace}
    \end{minipage}
    \caption{Example of traces used in our trace-driven evaluation}
    \label{fig:mahimahi_traces}
\end{figure}

\noindent\textbf{Trace collection:} 
We collected traces from WiFi and LTE networks, using \emph{saturatr}~\cite{winstein_stochastic_nodate}. In total, 20 network traces are collected, half of which are collected when the collector is stationary to the base station (LTE) or the Access Point (WiFi), and the other half are collected when the collector is moving at high speeds (i.e. in vehicles or on high-speed rails). In stationary scenarios, the network bandwidth is usually stable, while in high-mobility scenarios, the bandwidth fluctuates greatly. Examples of stationary and high-mobility traces are shown in Fig.~\ref{fig:mahimahi_traces}. The network delay and loss rate are also measured using \textit{ping}.


\begin{figure*}[htb]
    \centering
    \begin{minipage}{0.48\linewidth}
        \centering
        \includegraphics[width=0.7\linewidth]{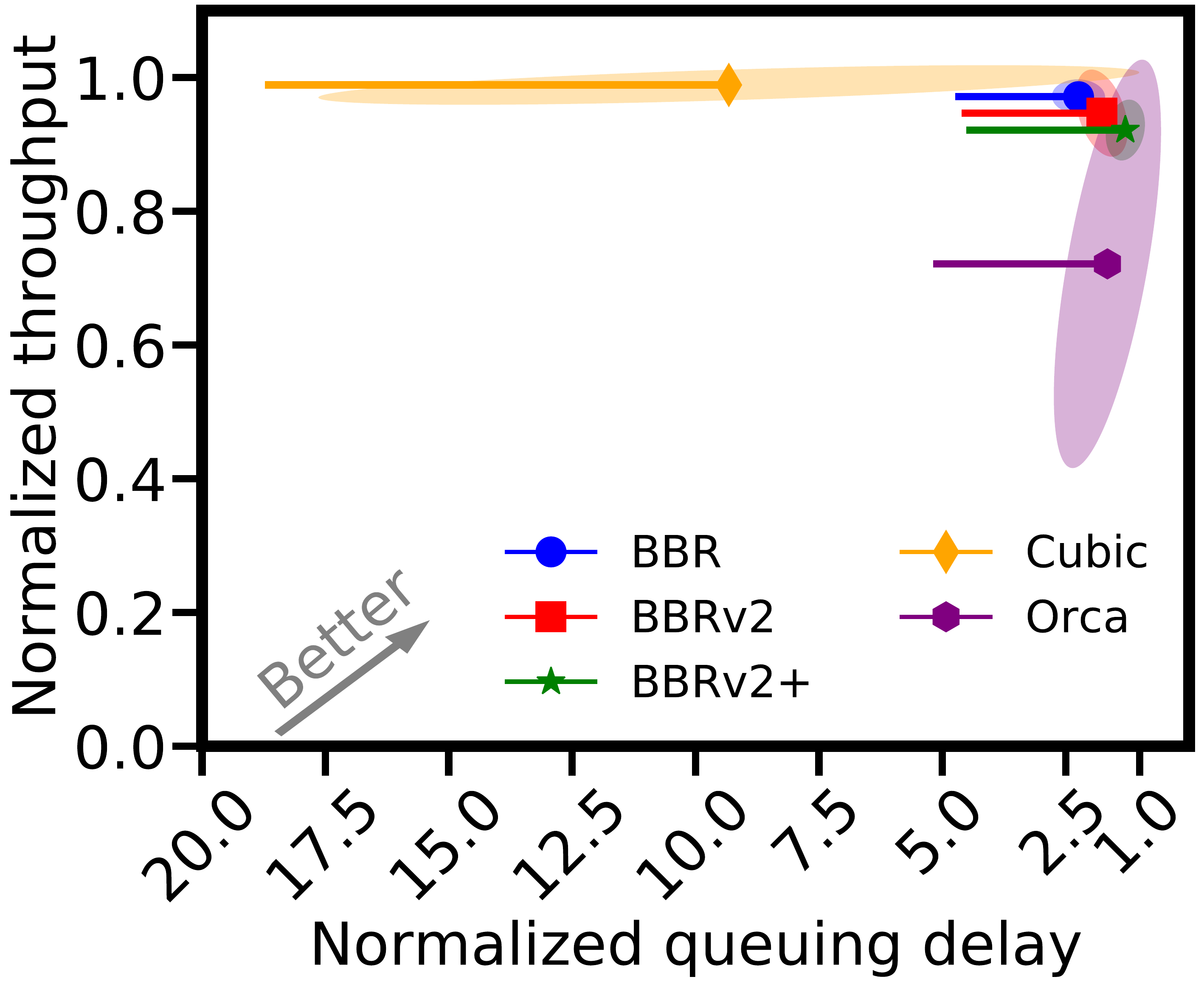}
        \subcaption{Stationary traces}
        \label{fig:mahimahi_static_tput_vs_delay}
    \end{minipage}
    \hfill
    \begin{minipage}{0.48\linewidth}
        \centering
        \includegraphics[width=0.7\linewidth]{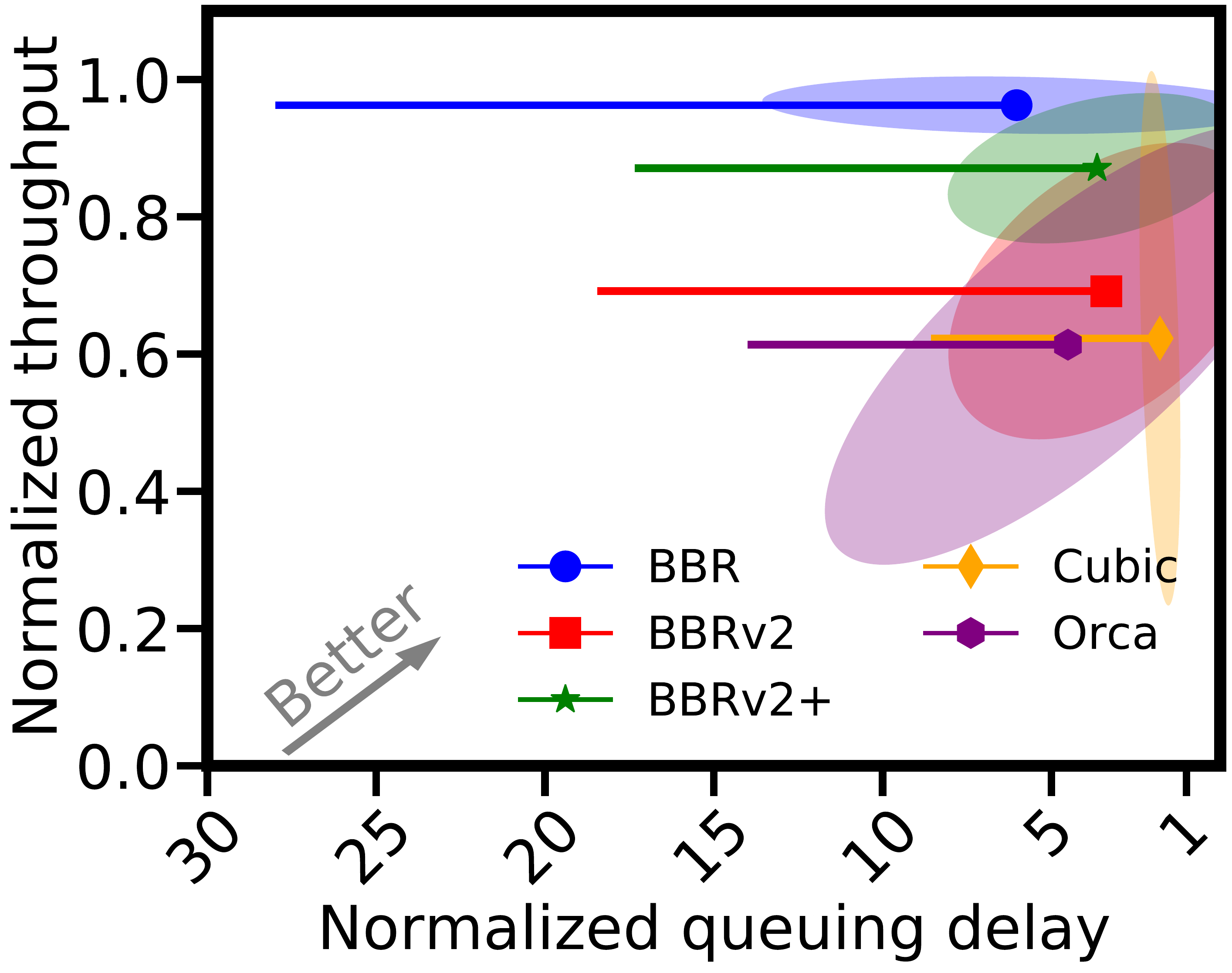}
        \subcaption{High-mobility traces}
        \label{fig:mahimahi_mobility_tput_vs_delay}
    \end{minipage}
    \caption{Normalized throughput and queuing delay of different CCAs. (markers: average throughput and queuing delay, left end of the lines: 95\%-tile of queuing delay, ellipses: the standard deviations of the throughput and queuing delay)}
    \label{fig:mahimahi_results}
\end{figure*}

\noindent\textbf{Experimental results:} The network buffer size is set to 1.5MB. The collected traces, including bandwidth dynamics, network delay, and loss rate are the inputs of the emulator. Using the same metrics as that in \S\ref{sec:eva:bw_change}, the results for the stationary and high-mobility scenarios are shown in Fig.~\ref{fig:mahimahi_static_tput_vs_delay} and Fig.~\ref{fig:mahimahi_mobility_tput_vs_delay} respectively.


In stationary scenarios, BBR, BBRv2, and BBRv2+ perform very close to each other because the bandwidth is usually stable. Cubic shows slightly better throughput for most of the time at the cost of high queuing delays. Orca has the lowest throughput probably because the network scenario where the model was trained is different from our collected traces, which also demonstrates the limitation of learning-based CCAs. 

In high-mobility scenarios, BBR and BBRv2+ achieve the highest and the second-highest throughput respectively. Meanwhile, BBRv2 and Cubic fail to achieve consistently high throughput across different high-mobility traces. Compared with BBRv2+, BBR achieves higher throughput at the cost that it incurs higher queuing delays as it is more aggressive. Orca fails to achieve consistently high throughput and low delays in high-mobility scenarios; the results of Orca raise a concern on the generalization ability of learning-based CCAs.

The above results of trace-driven emulation using Mahimahi demonstrate that BBRv2+ performs closely to BBR and BBRv2 in stationary network scenarios, but shows great improvements over BBRv2 in high-mobility scenarios as it has better responsiveness to bandwidth dynamics. 




\subsection{Summary of experimental results}

We can conclude from the above experiments that BBRv2+ succeeds to balance the aggressiveness of bandwidth probing and the fairness against loss-based CCAs. With such a balance, which is neither achieved by BBR nor BBRv2, BBRv2+ achieves higher throughput and lower delay than BBRv2 in scenarios where the bandwidth fluctuates, while keeping the advantages of BBRv2 with regard to inter-protocol fairness and reduced retransmissions under shallow buffers. Moreover, the dual-mode mechanism makes BBRv2+ able to co-exist with loss-based CCAs under deep buffers and the compensation mechanism for BDP estimation efficiently enhances the performance of BBRv2+ under high network jitters.


%% file: sections/relatedwork.tex
\label{sec:related_works}

\noindent\textbf{BBR evaluation:} Since BBR~\cite{Cardwell2016BBRCC} was released by Google in 2016, it has been examined under various network conditions by researchers~\cite{hock_experimental_2017,ma_fairness_2017,scholz_towards_2018,ware_modeling_2019,kumar_tcp_2019,cao_when_2019}. BBR is unfair when sharing a bottleneck link with Cubic flows or BBR flows with different RTTs~\cite{hock_experimental_2017,ma_fairness_2017,scholz_towards_2018,cao_when_2019}. In particular, BBR flows are always able to claim at least 35\% of the total bandwidth in deep-buffered networks when competing with Cubic flows~\cite{scholz_towards_2018}. This percentage, however, depends on the link capacity and delay, the bottleneck buffer size, and the number of BBR flows~\cite{ware_modeling_2019}. In shallow-buffered networks, BBR can lead to massive retransmissions~\cite{hock_experimental_2017}. Moreover, the throughput of BBR collapses when network jitters are high, either in experimental emulation~\cite{kumar_tcp_2019} or networks in high-speed train scenarios~\cite{wang_active-passive_2019}.

\noindent\textbf{BBR enhancement:} The issues of BBR identified by empirical studies motivated optimizations on BBR from various perspectives. For instance, \cite{kumar_tcp_2019, wang_active-passive_2019} proposed several modifications in the \textit{RTprop} estimation of BBR to counter against network jitters, \cite{ma_fairness_2017, yang_adaptive-bbr_2019, kim_delay-aware_2020} improved BBR's RTT fairness, \cite{zhang_modest_2018, song_bbr-cws_2020} improved BBR's inter-protocol fairness with loss-based CCAs, and \cite{mahmud_bbr-acd_2020} reduced BBR's aggressiveness in shallow-buffered networks to suppress the unnecessary retransmissions. 

\noindent\textbf{BBRv2 evaluation:} Google proposed BBRv2~\cite{cardwell_bbr_104} to solve the problems identified in BBR. In Google's early tests~\cite{cardwell_bbr_104, cardwell_bbr_105}, BBRv2 shows better inter-protocol fairness with loss-based CCAs and reduced retransmissions in shallow-buffered networks. There are also several evaluations on BBRv2~\cite{gomez_performance_2020, nandagiri_bbrvl_2020, song_understanding_2021, kfoury_emulation-based_2020}. Gomez et al.~\cite{gomez_performance_2020} and Nadagiri et al.~\cite{nandagiri_bbrvl_2020} studied the inter-protocol fairness and RTT fairness of BBRv2 through emulation. Kfoury et al.~\cite{kfoury_emulation-based_2020} evaluated BBRv2 in emulated networks and found that BBRv2 eliminates the massive retransmissions in shallow-buffered networks. Song et al.~\cite{song_understanding_2021} found that BBRv2 cannot quickly utilize the newly available link capacity when the bottleneck bandwidth increases. 

\vspace{0.5em}
Our work differs from the above studies in two perspectives: \textbf{(1)} from the measurement perspective, we not only systematically evaluate the performance of BBRv2 under various network conditions, but also analyze the reasons behind the observed performance issues; \textbf{(2)} from the optimization perspective, we propose BBRv2+ that addresses the shortcomings of BBRv2 while barely sacrificing its advantages in fairness and reduced retransmissions.

%% file: sections/conclusion.tex
\label{sec:conclusion}

In this paper, we first comprehensively evaluated BBRv2, revealed its pros and cons over BBR, and analyzed the reasons behind BBRv2's performance issues. Motivated by the results of BBRv2's evaluation, we propose BBRv2+ that incorporates \signame{} in its path model and redesigns the ProbeBW state to achieve a good balance between the aggressiveness of bandwidth probing and the fairness to loss-based CCAs. Extensive experiments demonstrate that BBRv2+ significantly improves the performance over BBRv2, especially in high-mobility scenarios, while barely sacrificing the advantages of BBRv2.